\documentclass[
reprint,
aps,
twocolumn,
superscriptaddress, 
floatfix,
]{revtex4}

\usepackage{gensymb}
\usepackage{amssymb}
\usepackage{bm, amsmath}
\usepackage{bm}
\usepackage{physics}
\usepackage[euler]{textgreek}
\usepackage{siunitx}
\usepackage[hidelinks]{hyperref}
\usepackage[version=4]{mhchem}
\usepackage{natbib}
\usepackage{balance}

\usepackage{graphicx}
\usepackage[caption=false]{subfig}
\usepackage[outdir=./]{epstopdf}

\usepackage[normalem]{ulem} 
\usepackage{verbatim}
\usepackage[draft]{todonotes}
\usepackage{float}

\begin{document}

\title{Impact of Dzyaloshinskii-Moriya and anisotropic exchange interactions on the cubic kagome antiferromagnets \ce{Mn3X} and \ce{Mn3AB}}

\author{J.~S.~R.~McCoombs}
\affiliation{Department of Physics and Atmospheric Science, Dalhousie University, Halifax, Nova Scotia, Canada B3H 3J5}
\author{A.~Zelenskiy}
\affiliation{Department of Physics and Atmospheric Science, Dalhousie University, Halifax, Nova Scotia, Canada B3H 3J5}
\author{M.~L.~Plumer}
\affiliation{Department of Physics and Atmospheric Science, Dalhousie University, Halifax, Nova Scotia, Canada B3H 3J5}
\affiliation{Department of Physics and Physical Oceanography, Memorial University of Newfoundland, St. John's, Newfoundland, Canada A1B 3X7}
\author{B.~W.~Southern}
\affiliation{Department of Physics and Astronomy, University of Manitoba, Winnipeg, Manitoba, Canada R3T 2N2}
\author{T.~L.~Monchesky}
\affiliation{Department of Physics and Atmospheric Science, Dalhousie University, Halifax, Nova Scotia, Canada B3H 3J5}
\date{\today}

\begin{abstract}
We perform a symmetry analysis of the ABC-stacked kagome planes of Mn atoms common to the $L1_2$ \ce{Mn3X} and antiperovskite \ce{Mn3AB} alloys. In addition to the single-ion-anisotropy and Kitaev-type anisotropic exchange known to stabilize \ang{120} spin structures in these materials, our analysis results in a staggered Dzyaloshinskii-Moriya interaction and further gamma-type anisotropic exchange between nearest-neighbor spins. The presence of these new terms is shown not to affect the energetics of the \ang{120} ground-states which explains their absence in prior minimal magnetic models. We go on to show, however, that their influence becomes apparent when spin-wave excitations are considered. We highlight these effects by calculating inelastic neutron scattering cross-sections to illustrate experimental means of probing the existence and relative strengths of these cloaked interactions.
\end{abstract}

\pacs{}

\maketitle

\section{Introduction}
Both the \ce{Cu3Au}-type \ce{Mn3X} alloys with X~$\in$~\{Rh, Pt, Ir\} and the antiperovskite \ce{Mn3AB} alloys with A~$\in$~\{Ga, Zn, Ni, Sn, Ag, Rh, Pt, \textit{etc}.\} and B~$\in$~\{C, N\} can exist in a cubic configuration belonging to the space group $Pm\bar{3}m$ with Mn ions occupying the face centers of the cubic unit cell, X/A atoms occupying the unit cell corners, and either a $B$ atom or a vacancy at the cell center (Fig.~\ref{fig:unit_cells}). In both structures, the octahedrally-arranged Mn ions form an ABC stacked array of 2D kagome layers which are contained to the \{111\}-family of planes. As is the case for the heavily studied 2D kagome system, the magnetic interactions in these 3D systems can be highly frustrated leading to a highly degenerate manifold of magnetic ground-states~\cite{Wills2001,Hemmati2012} when only nearest neighbor, isotropic exchange interactions are considered. By contrast, many of these \ce{Mn3X} and \ce{Mn3AB} compounds are found with an ordered $\vb{q}=0$, \ang{120} non-collinear N\'eel ordering~\cite{Krn1968} which is stabilized either by further-range exchange, anisotropic exchange, single-ion anisotropy (SIA), or a combination thereof.

In the \ce{Mn3AB} systems, the onset of \ang{120} ordering is usually associated with a dramatic, first-order increase in lattice dimension, though, with certain chemical doping procedures this transition can be smoothed out into a more continuous expansion~\cite{Takenaka2008,Takenaka2005,Matsuno2009,Iikubo2008}. The prospect of a controlled negative thermal expansion (NTE) has propelled interest in this class of materials. On the other hand, the \ce{Mn3X} alloys have been heavily studied in part as effective exchange-biasing layers in giant magnetoresistance (GMR) sensors~\cite{Tsunoda2010} with the ordered $L1_2$ \ce{Cu_3Au}-type structures enhancing the already large unidirectional magnetic anisotropy characteristic of disordered $\gamma$-type MnIr, MnRh, and MnPt alloys allowing for thinner pinning layers in GMR heterostructures.

In both the \ce{Mn3X} and \ce{Mn3AB} systems, non-collinear, antiferromagnetic spin arrangements have also garnered much interest recently due to the impact this type of spin-arrangement can have on various transport phenomenon. Chen \textit{et al.}~\cite{Chen2014} first predicted the possibility of a non-zero anomalous Hall effect (AHE) in the antiferromagnetic metal \ce{Mn3Ir} and the effect has since been experimentally realized in the related AB-stacked \ce{Mn3X} compounds \ce{Mn3Ge} and \ce{Mn3Sn}, the \ce{Cu3Au}-type compounds \ce{Mn3Ir} and \ce{Mn3Pt} as well as the antiperovskites \ce{Mn3GaN} and \ce{Mn3Ni_{1-x}Cu_xN}~\cite{Nayak2016,Nakatsuji2015,Iwaki2020,Liu2018,Hajiri2019,Zhao2019} solidifying the crucial symmetry breaking role these non-collinear spin arrangements play in enabling AHE even in the absence of a ferromagnetic moment. 

Non-collinear and particularly non-coplanar spin arrangements have also recently been studied for the impact that their magnetic excitations can have on thermal transport properties. The thermal Hall effect has been demonstrated in magnetic insulators with either non-coplanar spin arrangements or certain forms of the Dyzaloshinskii-Moriya (DM) interaction leading to dissipationless, transverse thermal current carried by spin-waves in an analogous fashion to the electronic Hall effect~\cite{Katsura2010,Owerre2017,Lu2019}. Though the materials we consider here are metals, our analysis remains valid for insulating magnetic structures with the same symmetries.

In this paper we derive a classical magnetic Hamiltonian for the ABC stacked kagome planes of Mn ions in both the \ce{Cu3Au}-type \ce{Mn3X} compounds and the antiperovskite \ce{Mn3AB} compounds based on symmetry considerations. Using this model we explore the space of possible ground-state configurations building upon the known $\vb{q}=0$ non-collinear configurations. We show that though the presence of the DM interaction in these systems does not alter the ground-state configurations in the $\vb{q}=0$ regime; its influence is activated when considering the magnonic excitations of these ground-states. We provide calculations which illustrate the influence the DM interaction has on the spin wave dispersion in these materials and demonstrate that the magnitude of the DM interaction should be measurable using inelastic neutron scattering.

\section{The symmetry-allowed magnetic Hamiltonian} \label{sec:symanal}
\begin{figure}[t]
	\centering
	\includegraphics[width=1.\columnwidth]{{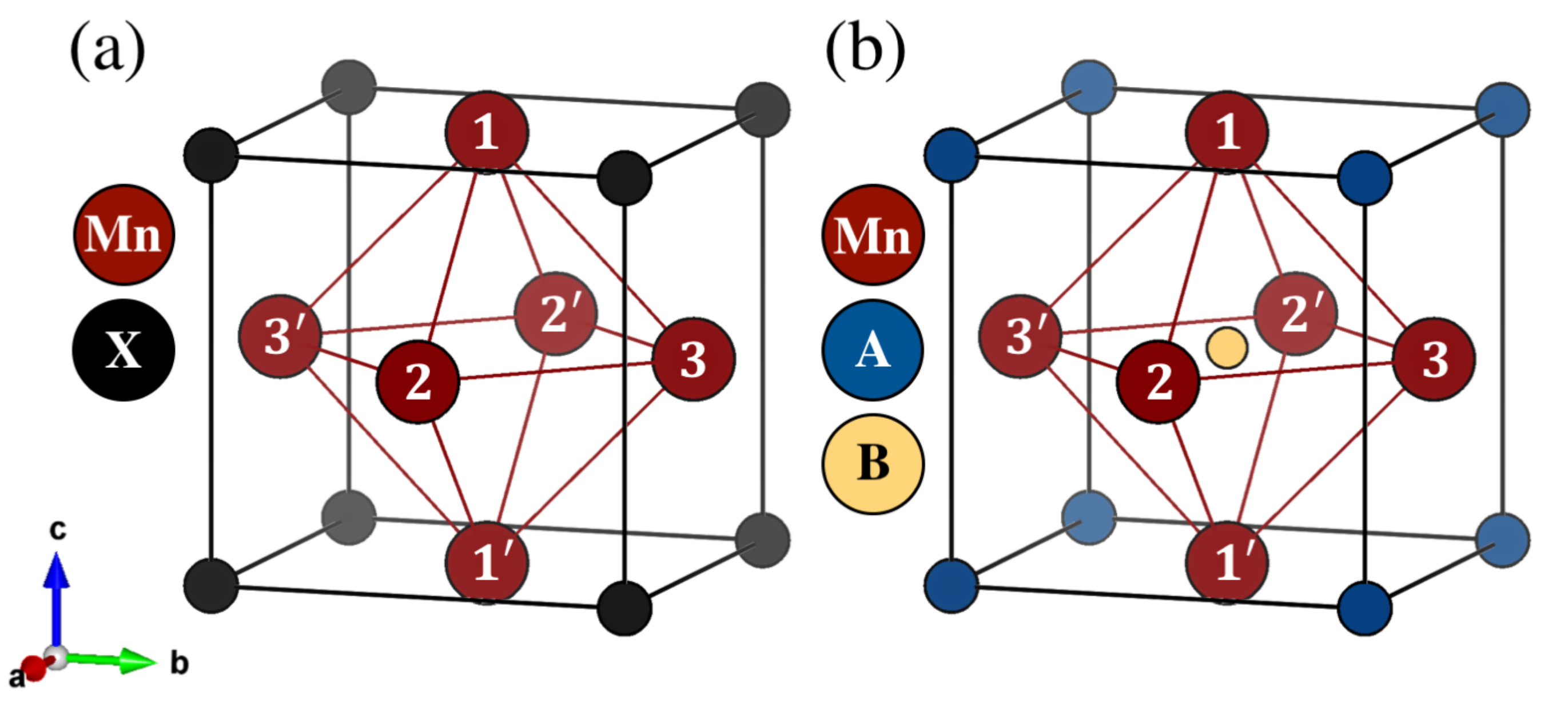}}
	\caption{The crystal structure of (a) the \ce{Cu3Au}-type \ce{Mn3X} compounds with the red Mn ions occupying the cubic faces and the black X atoms occupying the unit cell corners (b) the antiperovskite \ce{Mn3AB} compounds with the red Mn ions occupying the cubic faces, the black A atoms occupying the unit cell corners and the yellow B atom occupying the body-center of the unit cell.}
	\label{fig:unit_cells}
\end{figure}

For the purpose of this paper we consider only classical, bilinear, Heisenberg-type spin-spin interactions. We use the symmetries of the crystallographic space group to identify all allowable single-site and two-site interactions that have exchange paths contained within the first crystal unit cell. There are three Mn atoms in the unit cell shown in Fig.~\ref{fig:unit_cells}, which sit on three different sublattice positions. Each Mn atom has eight nearest neighbors, four from each of the other two sub-lattices. The spin-Hamiltonian can therefore be compactly expressed as 
\begin{equation}
	\mathcal{H} = \frac{1}{2}\sum_{\diamond}\sum_{ij}\vb{S}_i^T\mathcal{J}_{ij}\vb{S}_j
	\label{eq:genmagham}
\end{equation}
where the first sum runs over the octahedra (effectively the crystal unit cells), $\diamond$ and the second sum runs over the sub-lattice indices, $i,j \in (1,2,3,1',2',3')$. $\vb{S}$ is an 18-dimensional vector containing the six 3-dimensional spins that reside on the faces of the unit-cell,
\begin{equation}
	\vb{S}^T = \left(\vb{S}_1, \vb{S}_2, \vb{S}_3, \vb{S}_1^{\prime}, \vb{S}_2^{\prime}, \vb{S}_3^{\prime} \right)
	\label{eq:s18vec}
\end{equation}
and $\mathcal{J}$ is the $\left(18\times 18\right)$ coupling matrix
\begin{equation}
\mathcal{J} = \left[
\begin{matrix}
\mathcal{J}_{11} & \mathcal{J}_{12} & \mathcal{J}_{13} & \mathcal{J}_{11^{\prime}} & \mathcal{J}_{12^{\prime}} & \mathcal{J}_{13^{\prime}}\\[5pt]
\mathcal{J}_{21} & \mathcal{J}_{22} & \mathcal{J}_{23} & \mathcal{J}_{21^{\prime}} & \mathcal{J}_{22^{\prime}} & \mathcal{J}_{23^{\prime}}\\[5pt]
\mathcal{J}_{31} & \mathcal{J}_{32} & \mathcal{J}_{33} & \mathcal{J}_{31^{\prime}} & \mathcal{J}_{32^{\prime}} & \mathcal{J}_{33^{\prime}}\\[5pt]
\mathcal{J}_{1^{\prime}1} & \mathcal{J}_{1^{\prime}2} & \mathcal{J}_{1^{\prime}3} & \mathcal{J}_{1^{\prime}4} & \mathcal{J}_{1^{\prime}2^{\prime}} & \mathcal{J}_{1^{\prime}3^{\prime}}\\[5pt]
\mathcal{J}_{2^{\prime}1} & \mathcal{J}_{2^{\prime}2} & \mathcal{J}_{2^{\prime}3} & \mathcal{J}_{2^{\prime}1^{\prime}} & \mathcal{J}_{2^{\prime}2^{\prime}} & \mathcal{J}_{2^{\prime}3^{\prime}}\\[5pt]
\mathcal{J}_{3^{\prime}1} & \mathcal{J}_{3^{\prime}2} & \mathcal{J}_{3^{\prime}3} & \mathcal{J}_{3^{\prime}1^{\prime}} & \mathcal{J}_{3^{\prime}2^{\prime}} & J_{3^{\prime}3^{\prime}}
\end{matrix}
\right]
\label{eq:gencouplingmatrix}
\end{equation}
with 
\begin{equation}
\mathcal{J}_{ij} = \left[
\begin{matrix}
\mathcal{J}_{ij}^{xx} & \mathcal{J}_{ij}^{xy} & \mathcal{J}_{ij}^{xz} \\[5pt]
\mathcal{J}_{ij}^{yx} & \mathcal{J}_{ij}^{yy} & \mathcal{J}_{ij}^{yz}\\[5pt]
\mathcal{J}_{ij}^{zx} & \mathcal{J}_{ij}^{zy} & \mathcal{J}_{ij}^{zz}
\end{matrix}
\right].
\label{eq:gencouplingmatrix_element}
\end{equation}

\begin{figure}[t]
	\centering
	\includegraphics[width=1\columnwidth]{{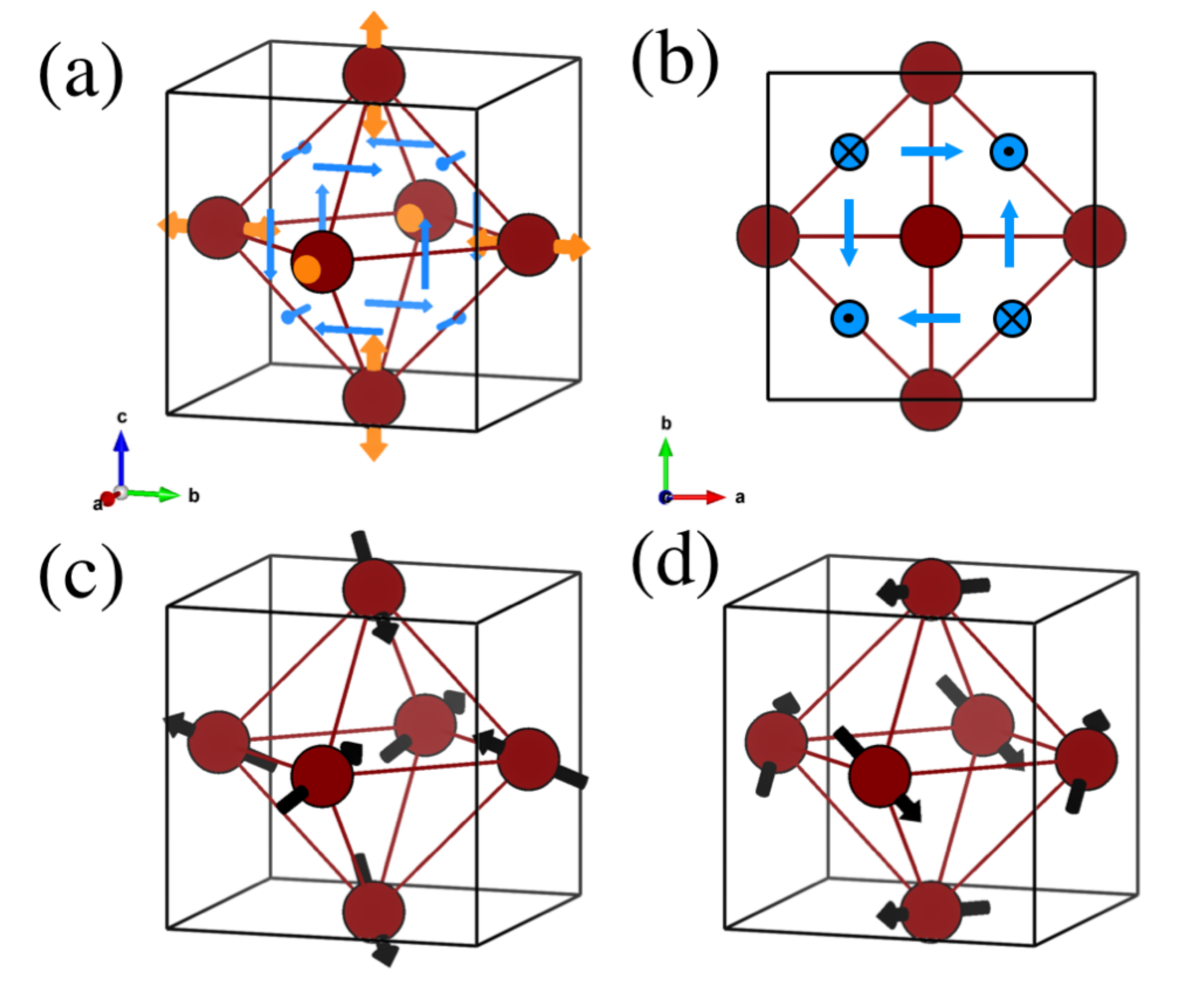}}
	\caption{(a) The magnetic Mn ions within the unit cell along with double-sided orange arrows indicating the easy/hard-axis directions on each of the three sites along with blue arrows indicating the direction of the DM vector at the bisector of each nearest-neighbor bond. (b) A top down view of the unit cell highlighting the directions of $\hat{\vb{D}}_{ij}$ in the top of the octahedron. (c) The T1 or $\Gamma_{4g}$ spin configuration. (d) The T2 or $\Gamma_{5g}$ spin configuration.}
	\label{fig:Mn3Ge_Paper_Easy_Axes}
\end{figure}

The coupling matrix, $\mathcal{J}$, contains all twelve nearest neighbor (NN) bonds shown in Fig.~\ref{fig:Mn3Ge_Paper_Easy_Axes}(a), all three single-site interactions, and all three next-nearest-neighbor (NNN) bonds. The factor of $\frac{1}{2}$ in front of $\mathcal{H}$ is due to the usual double-counting. 

By definition, the coupling matrix, $\mathcal{J}$, is equal to its transpose, thus, in general, it could contain 171 unique parameters. However, the spin-Hamiltonian must be invariant under all symmetry operations of the crystal, and so must be invariant under the generating operations of point group $O_h$. This restriction imposes five equations that $\mathcal{J}$ must obey, along with the condition that $\mathcal{J} = \mathcal{J}^T$. The remaining requirements are as follows:
\begin{equation}
	\begin{aligned}
		C_2 \mathcal{J} C_2^T &= \mathcal{J}, \\[5pt]
		C_2' \mathcal{J} C_2'^T &= \mathcal{J}, \\[5pt]
		C_3 \mathcal{J} C_3^T &= \mathcal{J}, \\[5pt]
		C_2'' \mathcal{J} C_2''^T &= \mathcal{J}, \\[5pt]
		\mathcal{I} \mathcal{J} \mathcal{I}^T &= \mathcal{J}.
	\end{aligned}
	\label{eq:symmetries}
\end{equation}
The operation $C_2$ refers to a $\pi$ rotation about the $[001]$ crystallographic direction, $C_2^{\prime}$ is a $\pi$ rotation about $[100]$, $C_3$ is a $2\pi/3$ rotation about $[111]$, $C_2^{\prime\prime}$ is a $\pi$ rotation about $[110]$ and $\mathcal{I}$ is the inversion operation. Upon application of these symmetries, the 171 free parameters contained in $\mathcal{J}$ are reduced to just eight. These are most conveniently displayed in the representative single-site $\mathcal{J}_{11}$, NN $\mathcal{J}_{23}$, and NNN $\mathcal{J}_{11^{\prime}}$ sub-matrices:
\begin{equation}
	\begin{aligned}
		\mathcal{J}_{11} &= \left[\begin{matrix}
		\mathcal{J}_{11}^{xx} & 0 & 0\\[5pt]
		0 & \mathcal{J}_{11}^{xx} & 0\\[5pt]
		0 & 0 & \mathcal{J}_{11}^{zz}\\[5pt]
		\end{matrix}\right], \\[5pt]
		\mathcal{J}_{23} &= \left[\begin{matrix}
		\mathcal{J}_{12}^{xx} & \mathcal{J}_{12}^{xy} & 0\\[5pt]
		\mathcal{J}_{12}^{yx} & \mathcal{J}_{12}^{xx} & 0\\[5pt]
		0 & 0 & \mathcal{J}_{12}^{zz}\\[5pt]
		\end{matrix}\right], \\[5pt]
		\mathcal{J}_{11^{\prime}} &= \left[\begin{matrix}
		\mathcal{J}_{11'}^{xx} & 0 & 0\\[5pt]
		0 & \mathcal{J}_{11'}^{xx} & 0\\[5pt]
		0 & 0 & \mathcal{J}_{11'}^{zz}\\[5pt]
		\end{matrix}\right],
	\end{aligned}
	\label{eq:samplecouplings}
\end{equation}
where all other $\mathcal{J}_{ij}$ can be obtained by an appropriate rotation. Following the decomposition procedure outlined in~\cite{Bertaut1963}, we can re-write the spin Hamiltonian in terms of more familiar bilinear coupling terms as follows:
\begin{align}
\mathcal{H} = \mathcal{H}_J + \mathcal{H}_D + \mathcal{H}_A + \mathcal{H}_B + \mathcal{H}_K + \mathcal{H}_{J^{\prime}}.
\label{eq:Compact_GenMagHam}
\end{align}
The various terms contributing to this general form of the Hamiltonian are summarized as:
\begin{equation}
\begin{aligned}
	\mathcal{H}_J &= \frac{J}{2}\sum_{\diamond}\sum_{ij}\vb{S}_i\vdot\vb{S}_j, \\
	\mathcal{H}_D &= \frac{D}{2}\sum_{\diamond}\sum_{ij}\vu{D}_{ij}\vdot\vb{S}_i\times\vb{S}_j, \\
	\mathcal{H}_A &= \frac{A}{2}\sum_{\diamond}\sum_{ij}\left(\vu{n}_{ij}\vdot \vb{S}_i\right)\left(\vu{n}_{ij}\vdot \vb{S}_j\right), \\
	\mathcal{H}_B &= \frac{B}{2}\sum_{\diamond}\sum_{ij}\left(\vu{l}_{ij}\vdot \vb{S}_i\right)\left(\vu{m}_{ij}\vdot \vb{S}_j\right) + \left(\vu{m}_{ij}\vdot \vb{S}_i\right)\left(\vu{l}_{ij}\vdot \vb{S}_j\right), \\
	\mathcal{H}_K &= \frac{K}{2}\sum_{\diamond}\sum_{i}\left(\vu{n}_i\vdot \vb{S}_i\right)^2,\\
	\mathcal{H}_{J^\prime} &= \frac{J^{\prime}}{2}\sum_{\diamond}\sum_{ii'}\vb{S}_i\vdot\vb{S}_i',
\end{aligned}
\label{eq:maghamterms}
\end{equation}
where the various unit-vectors, $\vu{D}_{ij}, \vu{n}_{ij}, \vu{m}_{ij}, \vu{l}_{ij},$ and $\vu{n}_{i}$ are defined in Eq.~\ref{eq:unit_vectors}.

This decomposition of the Hamiltonian via decomposition of the coupling matrix, $\mathcal{J}$, leads to terms representing NN isotropic exchange $(\mathcal{H}_J)$, DM interaction $(\mathcal{H}_D)$, Kitaev-type anisotropic exchange, $(\mathcal{H}_A)$, symmetric, off-diagonal, anisotropic exchange (sometimes referred to as Gamma-type exchange) $(\mathcal{H}_B)$, single-ion anisotropy $(\mathcal{H}_K)$ and NNN \textit{isotropic} exchange $\mathcal{H}_{J^{\prime}}$ where we have six free parameters, $J, J', D, A, B$ and $K$ (reduced from eight since one parameter amounts to an uninteresting constant energy shift and we have assumed for the scope of this paper that $A^{\prime}=0$, neglecting the second-order Kitaev-type term). These six free parameters of our model can be defined in terms of the eight in Eq.~\ref{eq:samplecouplings} as
\begin{equation}
	\begin{aligned}
		J &= \mathcal{J}_{12}^{xx} \\[5pt]
		J' &= \mathcal{J}_{11'}^{xx} = \mathcal{J}_{11'}^{zz}\\[5pt]
		D &= \frac{\mathcal{J}_{12}^{yx}-\mathcal{J}_{12}^{xy}}{2} \\[5pt]
		A &= \mathcal{J}_{12}^{zz} - \mathcal{J}_{12}^{xx} \\[5pt]
		B &= \frac{\mathcal{J}_{12}^{yx}+\mathcal{J}_{12}^{xy}}{2} \\[5pt]
		K &= \mathcal{J}_{11}^{zz} - \mathcal{J}_{11}^{xx}.
	\end{aligned}
\end{equation}
Of particular interest is the presence of the DM interaction between neighboring Mn ions. The revealed directions of $\vb{D}_{ij}$ through our symmetry analysis of $\mathcal{J}$ (shown in Fig.~\ref{fig:Mn3Ge_Paper_Easy_Axes}(a)) are consistent with Moriya's rules~\cite{Moriya1960}. They are also reminiscent of the form of the DM coupling in the isolated kagome plane geometry described by Elhajel \textit{et al.}~\cite{Elhajal2002} which included both an in-plane and out-of-plane component for the DM vectors. The presence of this in-plane component can be most easily justified by noting that the kagome planes are not themselves mirror planes, as is the case in the AB-stacked systems which consequently do not exhibit an in-plane component. Discussion of the DM interaction in both \ce{Mn3X} and \ce{Mn3AB} systems is almost completely lacking likely due to energetic cancellations in the most common T1 and $\Gamma_{5g}$ configurations as is pointed out by Chen \textit{et al.}~\cite{Chen2020}. However, as we will argue, the impact of this chiral term is not necessarily vanishing in these materials.
It is worth noting that the spin-structures of two representative examples from each of the classes, \ce{Mn_{3}X} and \ce{Mn_{3}AB}, can be explained rather simply using Eq.~\ref{eq:maghamterms}. The \ce{Cu3Au}-type structure \ce{Mn3Ir} was described by Szunyogh \textit{et al.}~\cite{Szunyogh2009} using a combination of $\mathcal{H}_{J}$, $\mathcal{H}_{A}$ and $\mathcal{H}_{K}$. The authors found that the Kitaev-type term and the SIA could effectively be rolled into a single, effective, easy-axis anisotropy term with a negative-valued $K_{\text{eff}}$ where the easy axis for each Mn atom corresponded to the normal direction of the cubic face it occupies as is shown in Fig.~\ref{fig:Mn3Ge_Paper_Easy_Axes}(a). The combination of easy-axis anisotropy and isotopic NN exchange of this form was investigated further by LeBlanc \textit{et al.}~\cite{LeBlanc2014} who showed that the ground-state spin structure of such a model should be the canted \ang{120} T1 (or $\Gamma_{4g}$) structure exhibiting a small net moment along one of its [111] directions. Figure \ref{fig:Mn3Ge_Paper_Easy_Axes}(c) shows the T1 structure with negligible canting. Likewise, the structure for the antiperovskite \ce{Mn3GaN} was described by Bertaut \textit{et al.}~\cite{Bertaut1968} using the subset of Eq.~\ref{eq:Compact_GenMagHam} including $\mathcal{H}_{J}$, $\mathcal{H}_{J'}$ and $\mathcal{H}_{K}$. In this case, the SIA axes shown in Fig.~\ref{fig:Mn3Ge_Paper_Easy_Axes} are \textit{hard} axes since the value for $K$ was determined to be positive. This energy penalty attributed to $S_{1z}$, $S_{2x}$ and $S_{3y}$ along with the ferromagnetic NNN coupling $(J'<0)$ leads to another \ang{120} structure in which each spin is confined to both the (111) plane and its respective cubic face which the authors dubbed the $\Gamma_{5g}$ configuration (also sometimes referred to as T2), as shown in Fig.~\ref{fig:Mn3Ge_Paper_Easy_Axes}(d). These two configurations will serve as representative examples of the two most commonly reported ground-state spin configurations in \ce{Mn3X} and \ce{Mn3AB} compounds with the goal of this paper being an expansion upon these simple examples of non-collinear order in \ce{Mn_{3}X} and \ce{Mn_{3}AB} systems with a particular emphasis on understanding the cloaked influence of the DM interactions in these materials.

It is important to point out that the coupling matrix $\mathcal{J}$ that was used to determine terms in the Hamiltonian contains $\left(3\times3\right)$ coupling matrices for each bond in the unit-cell. However, to simply describe spin-configurations in these systems, it is convenient to work in a three spin basis since each unit cell contains only three Mn atoms. This is most conveniently achieved by working with the Fourier transformed ${\mathcal{J}(\vb{q})}$, where each element is defined as
\begin{align}
\mathcal{J}_{ij}(\vb{q}) = \sum_{\vb{d}}\mathcal{J}_{ij}(\vb{d})e^{-i\vb{q}\cdot\vb{d}},
\label{eq:fouriertransformedcouplingmatrix}
\end{align}
with $\vb{d} = \vb{r}-\vb{r^{\prime}}$ being the vector which points from the Bravais lattice point of the unit cell containing first atom $i$ to the Bravais lattice point of the unit cell containing the second $j$. By doing this, we retain all the information contained in Eq. \ref{eq:gencouplingmatrix} while using a basis which conforms with the crystal structure. For example, in the case ${\vb{q}=0}$, we expect every Mn atom on sublattice 1 to point in a common direction and likewise for sublattices 2 and 3.

The Fourier transformed $\mathcal{J}(\vb{q})$ is as follows:
\begin{widetext}
	\begin{align}
	\mathcal{J}(\vb{q}) &= \left[\begin{matrix}
	\mathcal{J}_{11} + \mathcal{J}_{11^{\prime}}\gamma_{11^{\prime}}\left(\vb{q}\right) & \mathcal{J}_{12}\gamma_{12}\left(\vb{q}\right) + \mathcal{J}_{12^{\prime}}\gamma_{12^{\prime}}\left(\vb{q}\right) & \mathcal{J}_{13}\gamma_{13}\left(\vb{q}\right) + \mathcal{J}_{13^{\prime}}\gamma_{13^{\prime}}\left(\vb{q}\right)\\[6pt]
	~\mathcal{J}_{21}\gamma_{21}\left(\vb{q}\right) + \mathcal{J}_{21^{\prime}}\gamma_{21^{\prime}}\left(\vb{q}\right) & \mathcal{J}_{22} + \mathcal{J}_{22^{\prime}}\gamma_{22^{\prime}}\left(\vb{q}\right) & \mathcal{J}_{23}\gamma_{23}\left(\vb{q}\right) + \mathcal{J}_{23^{\prime}}\gamma_{23^{\prime}}\left(\vb{q}\right)~\\[6pt]
	\mathcal{J}_{31}\gamma_{31}\left(\vb{q}\right) + \mathcal{J}_{31^{\prime}}\gamma_{31^{\prime}}\left(\vb{q}\right) & \mathcal{J}_{32}\gamma_{32}\left(\vb{q}\right) + \mathcal{J}_{32^{\prime}}\gamma_{32^{\prime}}\left(\vb{q}\right) & \mathcal{J}_{33} + \mathcal{J}_{33^{\prime}}\gamma_{33^{\prime}}\left(\vb{q}\right)\\
	\end{matrix}\right],
\end{align}
\end{widetext}
where, for example, 
\begin{align}
\gamma_{12}\left(\vb{q}\right) = 1 + e^{i2\pi\left(h-l\right)},
\end{align}
with $\vb{q}=\left(2\pi h/a,2\pi k/a,2\pi l/a\right)$ and the $\mathcal{J}_{ij}$ are the same as in Eq.~\ref{eq:samplecouplings}.

We can now use the Fourier transformed $\mathcal{J}\left(\vb{q}\right)$ to gain insight into the periodicity of the ground-state spin structure for various combinations of the six free material parameters, $J,J^{\prime},D,A,B$ and $K$. We employ the Luttinger-Tisza method~\cite{Luttinger1946,Lyons1960} to minimize the Fourier-transformed Hamiltonian,
\begin{equation}
\begin{aligned}
\mathcal{H} = \frac{1}{2}\sum_{\vb{q}}\sum_{i,j}\vb{S}_i^T(\vb{q})\mathcal{J}_{ij}(\vb{q})\vb{S}_j(\vb{-q})
\end{aligned}
\label{eq:FT_ham}
\end{equation}
under the ``weak" constraint which states that the sum of the magnitudes of all spins must equal the number of spins. Under this weak constraint, minimization of (\ref{eq:FT_ham}) amounts to an eigenvalue problem in which the minimum energy eigenvalue of $\mathcal{J}(\vb{q})$, denoted $\lambda_0(\vb{q})$, corresponds to the ground-state energy for a given $\vb{q}$. We note that it is possible to obtain unphysical spin-configurations with non-unit-length spins while operating with the weak constraint. To validate a given solution, one must first check that each spin is unit-length.

\section{Minimum energy ordering vectors}

\begin{figure}[t]
	\centering
	\includegraphics[width=0.75\columnwidth]{{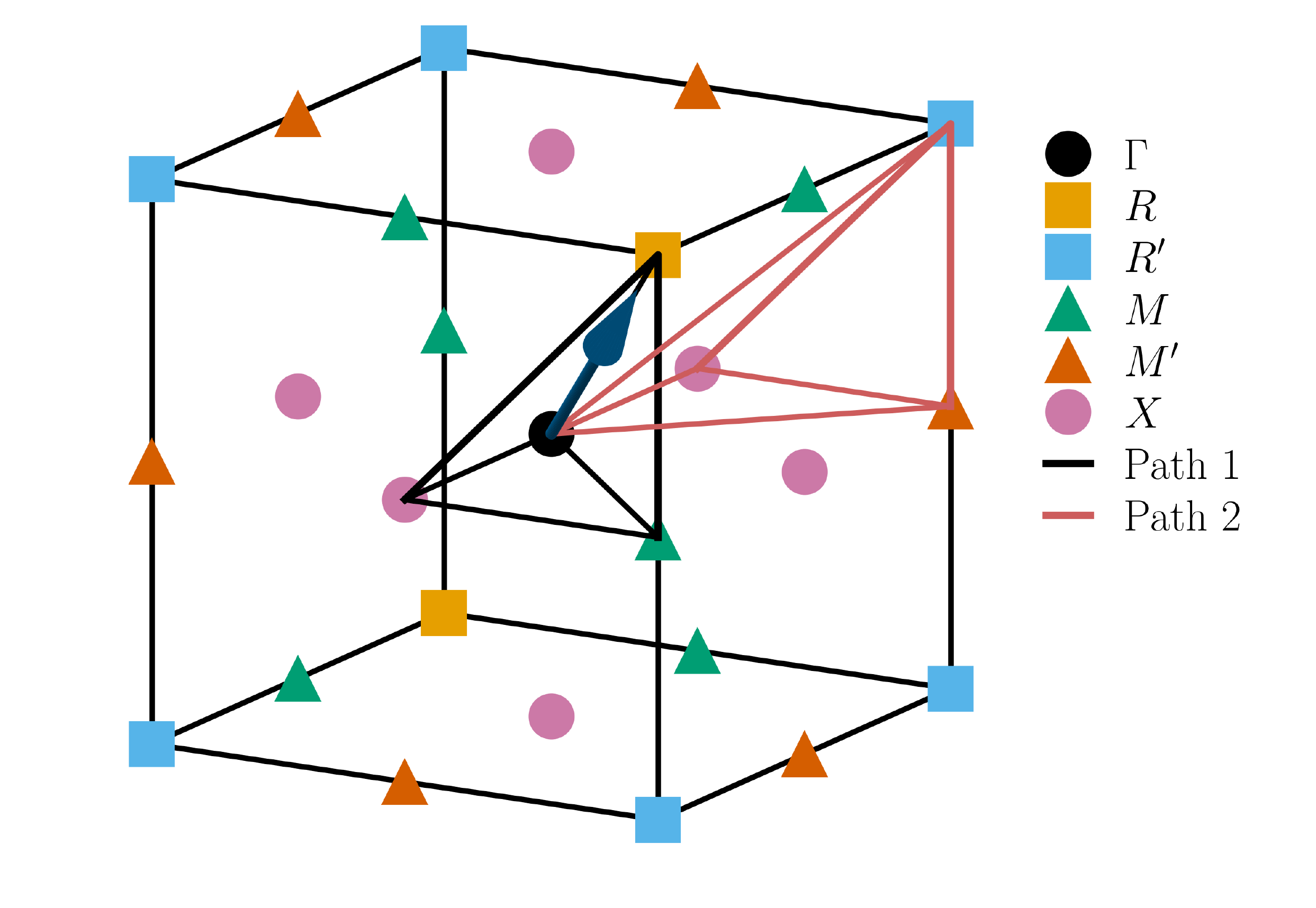}}
	\caption{The first Brillouin zone of the cubic system. The direction of the vector spin chirality is indicated by the arrow and is assumed to always be along the [111] direction without loss of generality.}
	\label{fig:BZ_path}
\end{figure}

\begin{figure}[t]
	\centering
	\includegraphics[width=1\columnwidth]{{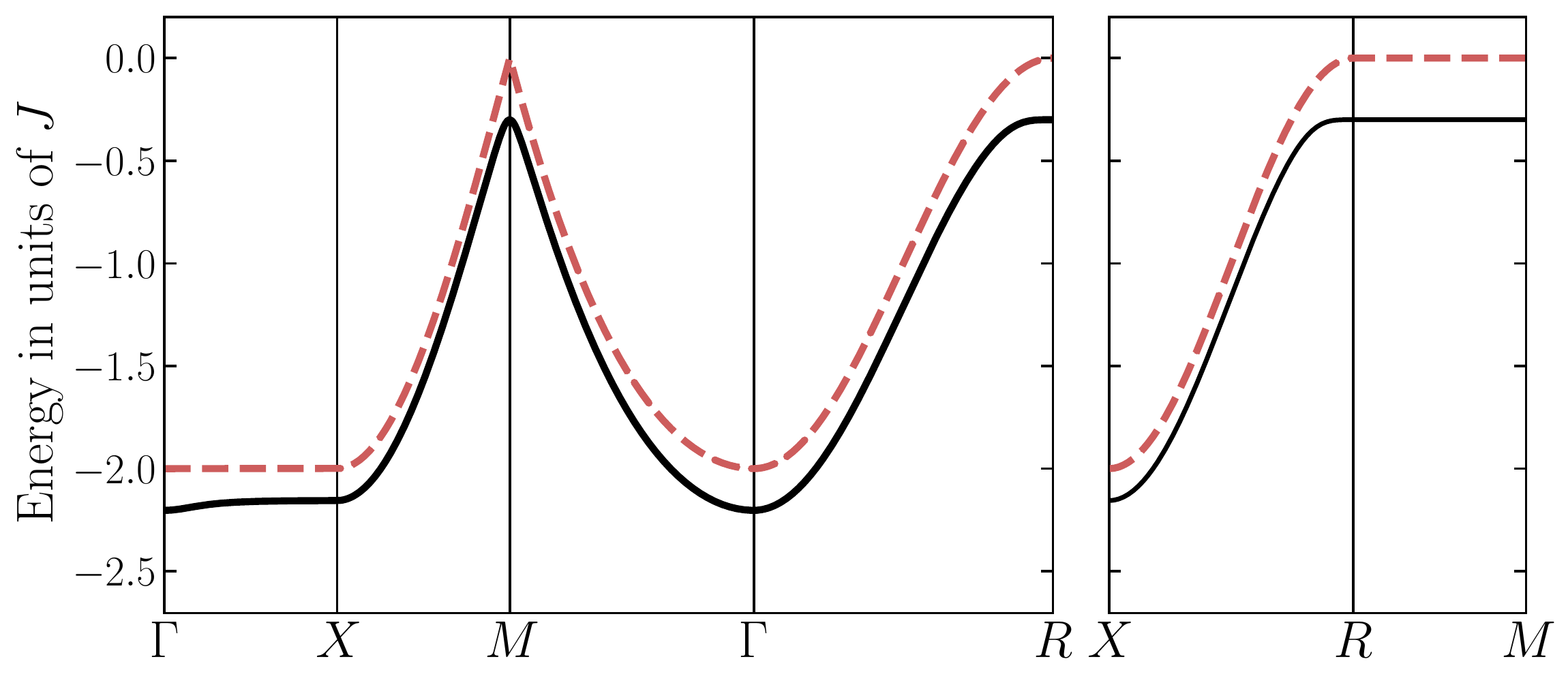}}
	\caption{(Solid black) The minimum energy eigenvalues $\lambda_0(\vb{q})$ for the coupling matrix $\mathcal{J}(\vb{q})$ with $J=1,K=0.2,$ and $J^{\prime}=-0.5$ and all other parameters set to zero. The clear minimum exists exclusively at the $\Gamma$-point, consistent with the $\vb{q}=0$ order known for this set of parameters. (Red dashed) The minimum energy eigenvalues $\lambda_0(\vb{q})$ for the coupling matrix $\mathcal{J}(\vb{q})$ with $J=1$ and all other parameters set to zero. $\lambda_{0}(\vb{q})$ is degenerate along the entire branch $\Gamma-X$ indicating a high degree of degeneracy.}
	\label{fig:T2g_NNN}
\end{figure}

\begin{figure}[t]
	\centering
	\includegraphics[width=0.9\columnwidth]{{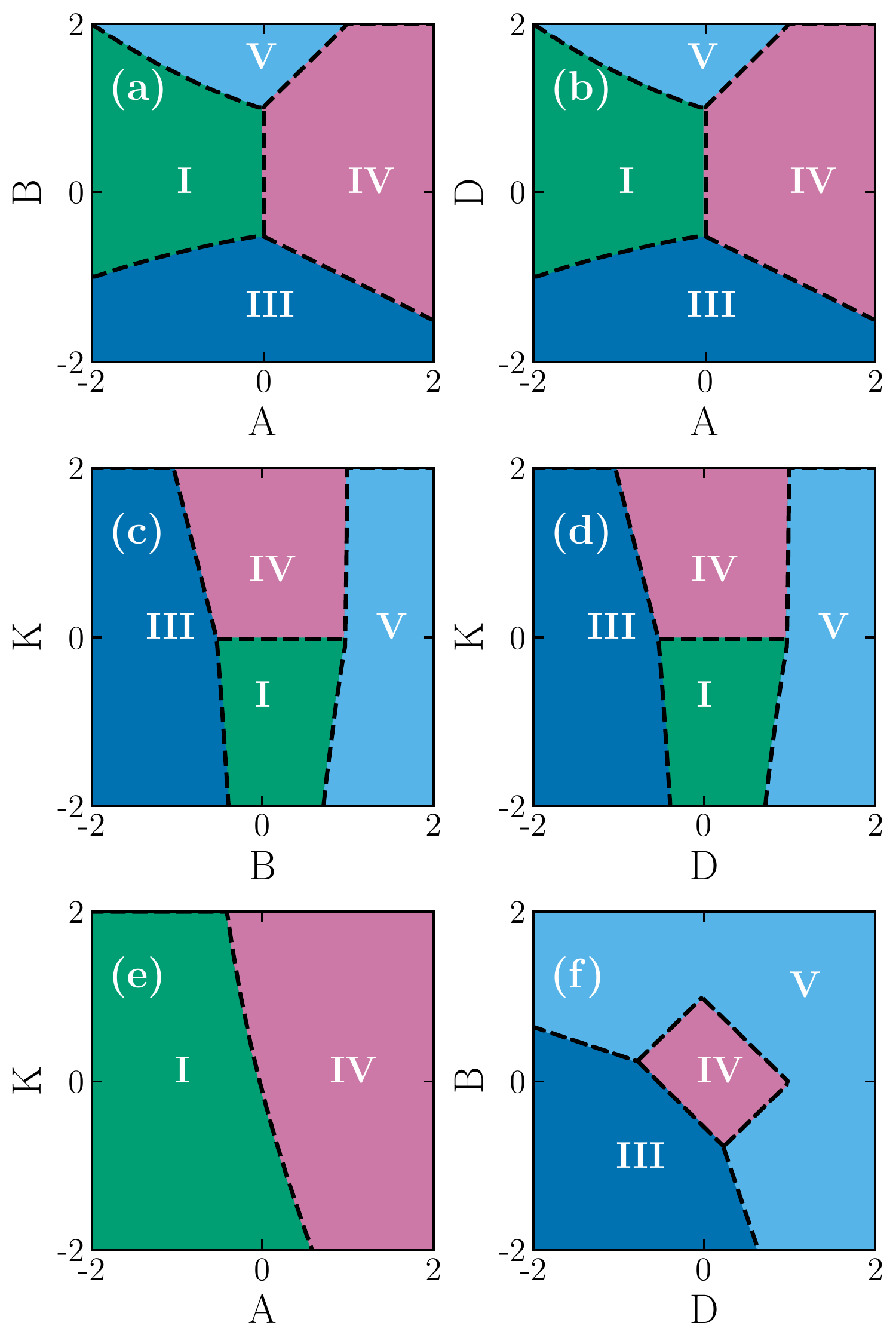}}
	\caption{Two dimensional slices through the four-dimensional sub-phase-space containing the parameters $A,B,D$ and $K$. The various regions identified correspond to one of the six possible energy-minimizing $\vb{q}$-vectors. Case I corresponds to a minimum at the $\Gamma$-point, II to minima at the $M$-points, III to minima at the $R$-points, IV to a macroscopic degeneracy between $\Gamma$ and $X$, V to a macroscopic degeneracy between $M$ and $R$ and VI to a macroscopic degeneracy over the entire reciprocal space path.}
	\label{fig:LT_Diagrams_pink}
\end{figure}

In practice, we expect the minimum energy state to have a wavevector that is along one of the high--symmetry directions of the lattice. Consequently, we evaluate the minimum energy eigenvalue $\lambda_0(\vb{q})$ along the path $\Gamma-X-M-\Gamma-R~||~X-R-M$ through the first Brillouin zone (Path 1 in Fig.~\ref{fig:BZ_path}).
As an example, consider the model described in section II for \ce{Mn3GaN} with $J=1$, $K=0.2$ and $J^{\prime}=-0.5$. The minimum energy $\lambda_0(\vb{q})$ of $\mathcal{J}(\vb{q})$ is plotted as the solid, black line in Fig.~\ref{fig:T2g_NNN} which demonstrates a clear energy minimum at the $\Gamma$-point. This $\vb{q}=0$ minimum energy state is consistent with the $\Gamma_{5g}$ ordering of this model described by Bertaut \textit{et al.} (though it is also consistent with the $\vb{q}=0$ T1 state and analysis of the eigenvectors is necessary to distinguish between the two (see section IV)). By contrast, if $K$ and $J^{\prime}$ are set to zero, $\lambda_{0}(\vb{q})$ becomes degenerate along the entire branch $\Gamma-X$ as can be seen in Fig.~\ref{fig:T2g_NNN} following the red, dashed line, indicating a highly degenerate ground-state manifold similar to previous studies~\cite{Harris1992,Chernyshev2015} in which so long as the angle between each spin in the system is \ang{120} the state qualifies as a viable magnetic ground-state.
We wish to study the extended model described by our general $\mathcal{J}(\vb{q})$ in the case of antiferromagnetic nearest-neighbor interactions, $J=1$. The minimum-energy $\vb{q}$ are summarized in the two-dimensional slices through configuration space shown in Fig.~\ref{fig:LT_Diagrams_pink} for this situation. There are a total of six different scenarios that can manifest. The different scenarios are labeled with the Roman numerals (I-VI). In cases I-III, the energy is minimized by a single $\vb{q}$ with $\vb{q}_{\text{I}}\in \Gamma$, $\vb{q}_{\text{II}}\in M$ and $\vb{q}_{\text{III}}\in R$. In cases IV-VI we find varying degrees of degeneracy throughout the BZ. Configuration IV has a flat ``dispersion" in the slice $\Gamma-X$, V has a similar behavior in the slice $M-R$ and VI has a degenerate $\lambda_0(\vb{q})$ for the whole BZ-path shown in Fig.~\ref{fig:BZ_path}. 

The phase boundaries between regions exhibit mixed degeneracy of the energy minimizing $\vb{q}$ on either side of the boundary indicative of the continuous transitions between regions. For example, at the phase boundary between regions IV and V, scenario VI is realized. Also worth noting is that a small, negative $J'$ breaks all degeneracies, with regions labeled IV switching to I and regions labeled V switching to II.
\section{Diagonalization of the $\vb{q}=0$ Hamiltonian}

We now explore the possible phases of our spin-system in the specific case $\vb{q}=0$ given the symmetry-allowed Hamiltonian found in section II which is shown in Eq.~(\ref{eq:qzeroJ}).
%
%
Importantly, all dependence on $D$ and $B$ has dropped out in the $\vb{q}=0$ limit. This is consistent with the two $\vb{q}=0$ configurations described in section II which had no DM interaction or Gamma-type exchange term present in their descriptions. In order to do this in a convenient and expedient manner we follow the method detailed by Essafi \textit{et al.}~\cite{Essafi2017}. Since the crystal structures of \ce{Cu3Au}-type \ce{Mn3X} and antiperovskite \ce{Mn3AB} belong to space group $Pm\bar{3}m$ with corresponding point-group symmetry $O_h$, the spin-Hamiltonian must also be invariant under all operations of $O_h$. These operations are detailed in Table~\ref{Tab:Characters} and Eq.~\ref{eq:representatives}. We have used this property once already to find the form of the coupling matrix, $\mathcal{J}$. Now we use it to aid in the diagonalization of $\mathcal{J}$.

\begin{widetext}
\begin{equation}
\mathcal{J}(\vb{q}=0) = \displaystyle \left[\begin{matrix}6 J^{\prime} & 0 & 0 & 4 J & 0 & 0 & 4 A + 4 J & 0 & 0\\0 & 6 J^{\prime} & 0 & 0 & 4 A + 4 J & 0 & 0 & 4 J & 0\\0 & 0 & 6 J^{\prime} + 2 K & 0 & 0 & 4 J & 0 & 0 & 4 J\\4 J & 0 & 0 & 6 J^{\prime} + 2 K & 0 & 0 & 4 J & 0 & 0\\0 & 4 A + 4 J & 0 & 0 & 6 J^{\prime} & 0 & 0 & 4 J & 0\\0 & 0 & 4 J & 0 & 0 & 6 J^{\prime} & 0 & 0 & 4 A + 4 J\\4 A + 4 J & 0 & 0 & 4 J & 0 & 0 & 6 J^{\prime} & 0 & 0\\0 & 4 J & 0 & 0 & 4 J & 0 & 0 & 6 J^{\prime} + 2 K & 0\\0 & 0 & 4 J & 0 & 0 & 4 A + 4 J & 0 & 0 & 6 J^{\prime}\end{matrix}\right].
\label{eq:qzeroJ}
\end{equation}
\end{widetext}

property once already to find the form of the coupling matrix, $\mathcal{J}$. Now we use it to aid in the diagonalization of $\mathcal{J}$.

The $\left(9\times9\right)$ symmetry operations, $\Gamma(g)$, are representative of the ten classes of $O_h$, where ${g\in\left(E,C_3,C_2,C_4,C_2^{\prime},i,S_4,S_6,\sigma_h,\sigma_d\right)}$. They are, however, a \textit{reducible} representation. By decomposition of the $\Gamma$ representation via a unitary transformation $\mathcal{U}^{\dagger}\Gamma(g)\mathcal{U}$, the $\Gamma$ representation can be decomposed into an irreducible representation. The transformed operations, $\mathcal{U}^{\dagger}\Gamma(g)\mathcal{U}$, will be block-diagonal, all with the same block structure. In fact, any $\left(9\times9\right)$ matrix which is invariant under $O_h$ will also be block-diagonalized by $\mathcal{U}$, including the coupling matrix of interest, $\mathcal{J}$.

Our first task is to find the relevant characters of the $\Gamma$ representation operators, $\chi(g)$. These characters are simply the trace of one representative operation from each class, $g$. To simplify matters, we note that the full operation $\Gamma(g)$ is the tensor product of the corresponding spin permutation matrix $\mathcal{P}(g)$ and the spin rotation matrix $\mathcal{R}(g)$:
\begin{align}
\Gamma(g) = \mathcal{P}(g)\otimes\mathcal{R}(g),
\end{align}
thus the corresponding character is just the product of the permutation operation's character and the rotation operation's character:
\begin{align}
\chi(g) = \chi_\mathcal{P}(g)\chi_\mathcal{R}(g).
\end{align}
\begin{table}[H]
	\centering
	\begin{ruledtabular}
	\begin{tabular}{c|ccc}
		Operation          & $\chi_R$ & $\chi_P$ & $\chi$ \\ \hline
		$C_3: \{3^+_{111}\}$    & 0                   & 0                   & 0                   \\
		$C_2: \{2_{110}\}$      & -1                  & 0                   & 0                   \\
		$C_4: \{4^+_{001}\}$    & 1                   & 2                   & 2                   \\
		$C_2': \{2_{001}\}$     & -1                  & 2                   & -2                  \\
		$\mathcal{I}: \{-1\}$               & 3                   & 0                   & 0                   \\
		$S_4: \{-4^+_{001}\}$   & 1                   & 0                   & 0                   \\
		$S_6: \{-3^+_{111}\}$   & 0                   & 0                   & 0                   \\
		$\sigma_h: \{m_{001}\}$ & -1                  & 4                   & -4                  \\
		$\sigma_d: \{m_{110}\}$  & -1                  & 2                   & -2                 
	\end{tabular}%
	\end{ruledtabular}
\caption{Representative symmetry operations from each class of $O_h$ with their corresponding permutation characters $\chi_P$, rotation characters $\chi_R$ and total characters $\chi$.}
\label{Tab:Characters}
\end{table}
We can now use these characters to find a decomposition, 
\begin{align}
\Gamma = \oplus_\text{I} \gamma_\text{I} \Gamma_\text{I}
\end{align}
of the spin configuration into its irreducible representations where the coefficients, $\gamma_I$ are
\begin{align}
\gamma_\text{I} = \frac{1}{n}\sum_{g\in O_h} \chi_\text{I}(g)\chi(g)
\end{align}
where $n$ is is the total number of symmetry elements, in this case $n=48$, and $\chi_\text{I}(g)$ are the characters summarized in Table~\ref{Tab:CharacterTable}. This results in the following decomposition for the 9-D representation:
\begin{align}
\Gamma = 2T_{1g} \oplus T_{2g}.
\end{align}
As a result, we expect that our coupling matrix should be block-diagonalizable into three three-dimensional blocks, two which transform as $T_{1g}$ and one which transforms as $T_{2g}$.
%
%
It is possible to describe 9 basis vector configurations that transform as the representations $T_{1g}$ and $T_{2g}$. These nine symmetry-adapted-order-parameters (SAOP's) will be convenient for interpreting the ground-state spin configuration of the $\vb{q}=0$ spin-Hamiltonian. They are defined as follows in the $\vb{S}=\left[\vb{S}_{1},\vb{S}_{2},\vb{S}_{3}\right]$ basis:
\begin{equation}
	\begin{aligned}
	\vb{T}_{1g}^{Fx} &= \left[100,100,100\right]\\[5pt]
	\vb{T}_{1g}^{Fy} &= \left[010,010,010\right]\\[5pt]
	\vb{T}_{1g}^{Fz} &= \left[001,001,001\right]\\[5pt]
	\vb{T}_{1g}^{Ax} &= \frac{1}{\sqrt{2}}\left[100,-200,100\right]\\[5pt]
	\vb{T}_{1g}^{Ay} &= \frac{1}{\sqrt{2}}\left[010,010,0-20\right]\\[5pt]
	\vb{T}_{1g}^{Az} &= \frac{1}{\sqrt{2}}\left[00-2,001,001\right]\\[5pt]
	\vb{T}_{2g}^{x} &= \sqrt{\frac{3}{2}}\left[100,000,-100\right]\\[5pt]
	\vb{T}_{2g}^{y} &= \sqrt{\frac{3}{2}}\left[010,0-10,000\right]\\[5pt]
	\vb{T}_{2g}^{z} &= \sqrt{\frac{3}{2}}\left[000,001,00-1\right].\\
	\end{aligned}
\end{equation}
Whereas configurations from the $T_{1g}^{F\alpha}$ block conform to the strong constraint with unit-length spins, configurations from the $T_{1g}^{A\alpha}$ and $T_{2g}$ blocks do not. To realize physical spin-configurations in these blocks, linear combinations must be formed. Two example physical configurations are as follows,
\begin{equation}
\begin{aligned}
\vb{T}_{1g}^{A} &= \frac{1}{\sqrt{3}}\left(\vb{T}_{1g}^{Ax} + \vb{T}_{1g}^{Ay} + \vb{T}_{1g}^{Az}\right)\\
\vb{T}_{2g} &= \frac{1}{\sqrt{3}}\left(-\vb{T}_{2g}^{x} + \vb{T}_{2g}^{y} - \vb{T}_{2g}^{z}\right)\\
\end{aligned}
\end{equation}
and are shown in Fig.~\ref{fig:Mn3Ge_Paper_Easy_Axes}(c,d) where it becomes obvious that the $T_{1g}^{A}$ configuration corresponds to T1($\Gamma_{4g}$) ordering while $T_{2g}$ corresponds to T2($\Gamma_{5g}$). Rewriting the coupling matrix, $\mathcal{J}(\vb{q}=0)$ in the SAOP basis reveals the utility in working in such a basis. The transformed $\mathcal{J}$ is 
\begin{equation}
\mathcal{J^{\prime}}(\vb{q}=0) = \left[
\begin{matrix}
P & 0 & 0 & S & 0 & 0 & 0 & 0 & 0\\
0 & P & 0 & 0 & S & 0 & 0 & 0 & 0\\
0 & 0 & P & 0 & 0 & S & 0 & 0 & 0\\
S & 0 & 0 & Q & 0 & 0 & 0 & 0 & 0\\
0 & S & 0 & 0 & Q & 0 & 0 & 0 & 0\\
0 & 0 & S & 0 & 0 & Q & 0 & 0 & 0\\
0 & 0 & 0 & 0 & 0 & 0 & R & 0 & 0\\
0 & 0 & 0 & 0 & 0 & 0 & 0 & R & 0\\
0 & 0 & 0 & 0 & 0 & 0 & 0 & 0 & R\\
\end{matrix}
\right]
\end{equation}
with 
\begin{equation}
\begin{aligned}
P &= \frac{8A}{3} + 8J + 6J^{\prime} + \frac{2K}{3}\\[5pt]
Q &= \frac{4A}{3} - 4J + 6J^{\prime} + \frac{4K}{3}\\[5pt]
R &= -4A - 4J + 6J^{\prime}\\[5pt]
S &= \frac{2\sqrt{2}}{3}\left(2A-K\right).
\end{aligned}
\end{equation}
This form of the coupling matrix is nearly diagonal, and one can gain insight as to which eigenvectors, or SAOP's, will need to mix in order to create the minimal energy state given a prescribed set of $J,A,J^{\prime},$ and $K$. Namely, in the case where $A=K=0$, we can see immediately that for $J<0$ the ground state belongs to the $T_{1g}^{F}$ family with some $SO(3)$ global rotational symmetry as is the case in the classical 3D Heisenberg ferromagnet. Meanwhile, if $J>0$, we find that the $T_{1g}^{A}$ and $T_{2g}$ antiferromagnetic states minimize the energy and are degenerate with no mixing between $T_{1g}^A$ and $T_{1g}^F$ since the off diagonal terms are zero. If $A$ and/or $K$ are \textit{not} zero, we expect these two modes to intermix. To move forward, we must take one more step to remove the mixing between our mathematically intuitive but physically arbitrary SAOP's in the $T_{1g}^{F}$ and $T_{1g}^A$ families. To do this we create six new basis vectors which are simply linear combinations of $\vb{T}_{1g}^{F\alpha}$ and $\vb{T}_{1g}^{A\alpha}$:
\begin{equation}
\begin{aligned}
\vb{T}_{1g}^{\text{I}\alpha} &= \frac{\xi \vb{T}_{1g}^{F\alpha} + \vb{T}_{1g}^{A\alpha}}{\sqrt{1+\xi^2}}\\[5pt]
\vb{T}_{1g}^{\text{II}\alpha} &=  \frac{\vb{T}_{1g}^{F\alpha} + \zeta\vb{T}_{1g}^{A\alpha}}{\sqrt{1+\zeta^2}}.
\end{aligned}
\end{equation}
where the constants $\xi$ and $\zeta$ are defined in Eq.~(\ref{eq:constants}). With these slightly more cumbersome order parameters defined, the $\vb{q}=0$ Hamiltonian now becomes fully diagonal with the coupling matrix re-defined as
\begin{equation}
\mathcal{J^{\prime\prime}}(\vb{q}=0) = \left[
\begin{matrix}
P' & 0 & 0 & 0 & 0 & 0 & 0 & 0 & 0\\
0 & P' & 0 & 0 & 0 & 0 & 0 & 0 & 0\\
0 & 0 & P' & 0 & 0 & 0 & 0 & 0 & 0\\
0 & 0 & 0 & Q' & 0 & 0 & 0 & 0 & 0\\
0 & 0 & 0 & 0 & Q' & 0 & 0 & 0 & 0\\
0 & 0 & 0 & 0 & 0 & Q' & 0 & 0 & 0\\
0 & 0 & 0 & 0 & 0 & 0 & R & 0 & 0\\
0 & 0 & 0 & 0 & 0 & 0 & 0 & R & 0\\
0 & 0 & 0 & 0 & 0 & 0 & 0 & 0 & R\\
\end{matrix}
\right],
\end{equation}
with $P'$ and $Q'$ defined in Eq.~\ref{eq:constants} and the three ${(3\times3)}$ blocks now made very apparent. Thus, whichever of $P'$, $Q'$ or $R$ is minimized for a given set of $J, A, J', K$ will saturate the spin-configuration in its corresponding normalized linear combination of SAOP's.

\begin{figure}[!h]
	\centering
	\includegraphics[width=0.85\columnwidth]{{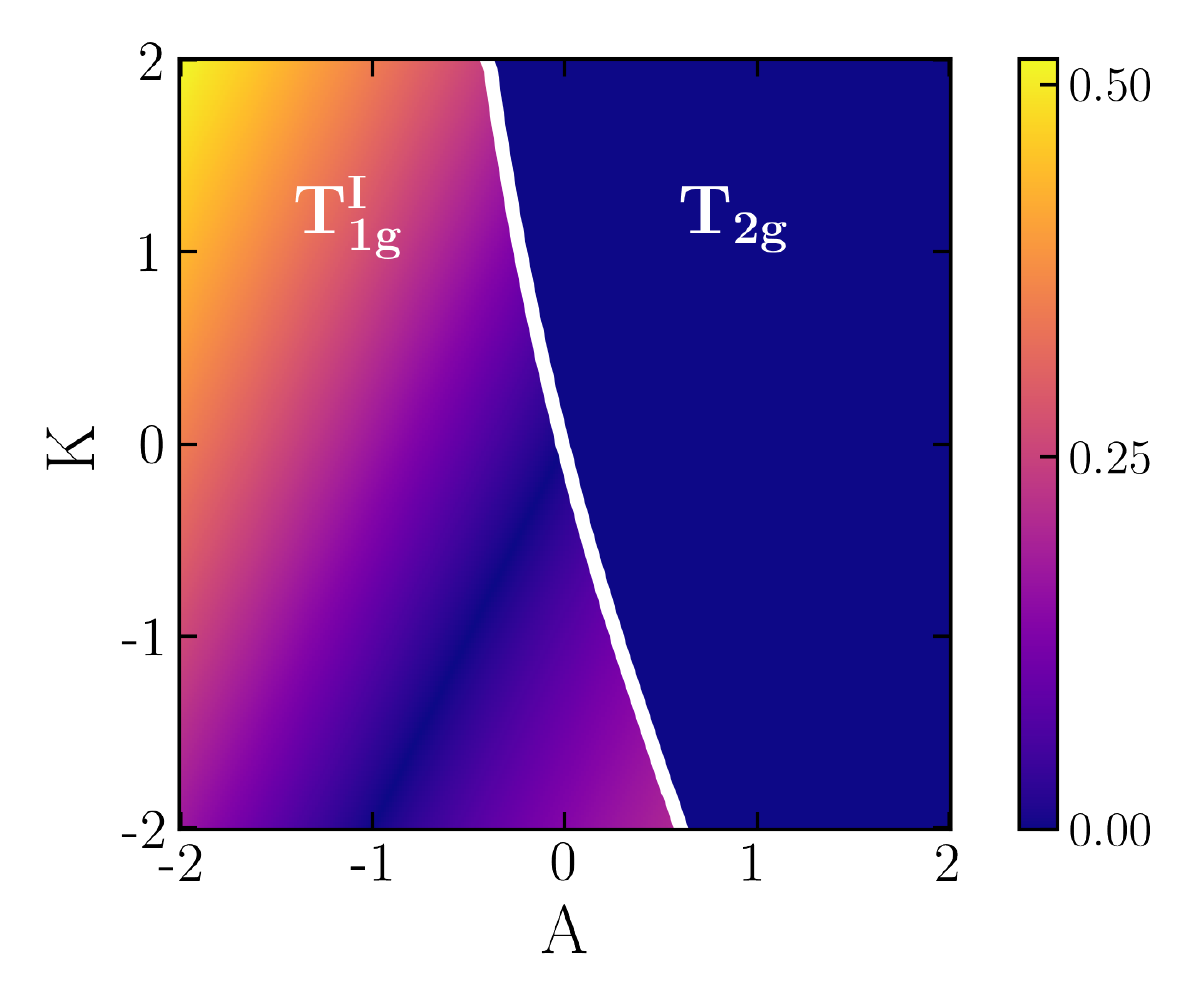}}
	\caption{Phase diagram for the parameters $A$ and $K$ with $J=1$ and $J^{\prime}=-0.1$. The intensity in the region labeled $T_{1g}$ corresponds to the weight of the $T_{1g}^F$ component in the linear combination $T_{1g}^I$ which is associated with the amount of ferromagnetic canting toward the [111] direction. The region labeled $T_{2g}$ has no ferromagnetic contribution, it couples to neither of the $T_{1g}$ modes.}
	\label{fig:q0_phase_diagram}
\end{figure}

Returning to the two examples in Section II, we can immediately explain the canted T1 ordering for \ce{Mn3Ir} since for ${J=1, K<0, A=0, J'=0}$, the minimum energy eigenvalue is $P'$ which means the spin configuration should be $\vb{T}_{1g}^{\text{I}}$ ($T1$ plus canting toward the $[111]$ direction). Likewise, for \ce{Mn3GaN} with ${J=1,K>0,A=0,J'<0}$, we immediately find the minimum energy eigenvalue to be $R$ which corresponds to the $\vb{T}_{2g}$ ($\Gamma_{5g}$) configuration.
\begin{widetext}
\begin{equation}
\begin{aligned}
P' &= 2 A + 2 J + 6 J^{\prime} + K - \sqrt{4 A^{2} + 8 A J - 4 A K + 36 J^{2} - 4 J K + K^{2}}\\[5pt]
Q' &= 2 A + 2 J + 6 J^{\prime} + K + \sqrt{4 A^{2} + 8 A J - 4 A K + 36 J^{2} - 4 J K + K^{2}}\\[5pt]
\xi &= \frac{2 \sqrt{2} \left(- 2 A + K\right)}{2 A + 18 J - K + 3 \sqrt{4 A^{2} + 8 A J - 4 A K + 36 J^{2} - 4 J K + K^{2}}}\\[5pt]
\zeta &= \frac{2 A + 18 J - K - 3 \sqrt{4 A^{2} + 8 A J - 4 A K + 36 J^{2} - 4 J K + K^{2}}}{2 \sqrt{2} \left(- 2 A + K\right)}.
\label{eq:constants}
\end{aligned}
\end{equation}
\end{widetext}
Using this methodology, we can take our analysis one step further to examine the amounts of $T_{1g}^{F}$ and $T_{1g}^{A}$ that make up our newly defined $T_{1g}^I$ state. Figure~\ref{fig:q0_phase_diagram} shows a similar picture to Fig.~\ref{fig:LT_Diagrams_pink}(e) with the added insight of just how much canting along the [111] direction is induced for various values of $A$ and $K$. Importantly, this is also a measure of the scalar spin chirality, $\chi_{ijk} = \vb{S}_i\cdot\left(\vb{S}_j\times\vb{S}_k\right)$ of the configuration. Notably, for nearly the entire portion of the phase diagram corresponding to the $T_{1g}$ configuration, $\chi_{ijk}$ is non-zero due to coupling with the $\vb{T}_{1g}^F$ mode inducing canting toward the [111] direction while for the $T_{2g}$ region, $\chi_{ijk}$ is identically zero since no coupling with the ferromagnetic mode is permitted, resulting in a strictly coplanar arrangement. In both cases the vector spin chirality, $\vb*{\kappa}_{ijk} \propto \vb{S}_i\times\vb{S}_j + \vb{S}_j\times\vb{S}_k + \vb{S}_k\times\vb{S}_i$ is directed along the [111] crystallographic direction, reflecting the shared handedness of the two configurations.

\section{Influence on spin-wave dispersion}

As was made clear in section IV, the presence of the DM interaction and gamma-type anisotopic exchange have no impact on the ground-state spin configurations in the $\vb{q}=0$ regime. The energetic contributions due to these terms cancel out identically in these ferromagnetic sub-lattice configurations. However, as we will show, one should still be able to glean information on their presence and and even qualitative measurements of their magnitudes by considering the elementary magnetic excitations of the ${\vb{q}=0}$ ground-states.

Following the methods outlined in Ref.~\cite{Toth2015} and Ref.~\cite{Petit2011}, we calculate first the spin wave dispersion along the paths outlined in Fig.~\ref{fig:BZ_path}. It is important to note that we consider here a system which has spontaneously aligned in such a way that the vector-spin chirality (or the magnetic moment for $T_{1g}$ with $K<0$) is aligned solely with the [111] crystallographic direction. This alignment breaks some of the crystallographic symmetries and as a result, the point-group symmetry of the system is reduced to $D_{3d}$ for coplanar arrangements and $S_6$ for non-coplanar arrangements. Consequently, it is necessary to probe both paths through the first cubic Brillouin zone indicated in Fig.~\ref{fig:BZ_path} since the points labeled $M'$, for example, indicate wavevectors that lie within the chosen kagome planes, perpendicular to $\vec{\kappa}$, while points labeled $M$ indicate a wavevector with components both perpendicular and parallel to $\vec{\kappa}$.

\begin{figure}[t]
	\centering
	\includegraphics[width=1.0\columnwidth]{{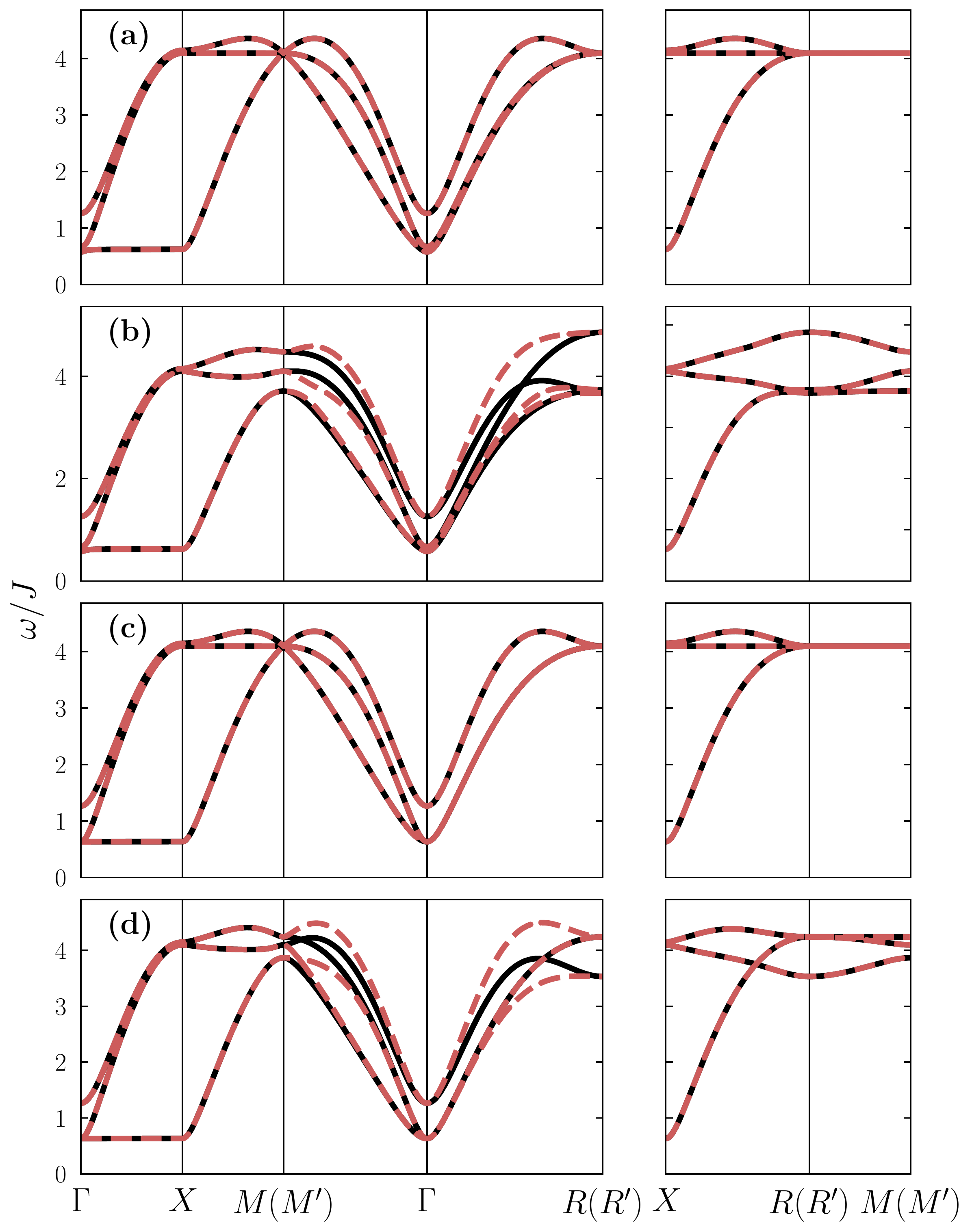}}
	\caption{Spin-wave dispersion for (a) the $T_{1g}$ ground-state with $K=-0.1$, $D=0$, (b) the $T_{1g}$ ground-state with $K=-0.1$ and $D=-0.2$, (c) the $T_{2g}$ ground-state with $K=0.1$, $D=0$, and (d) the $T_{2g}$ ground-state with $K=0.1$, $D=-0.2$. In all cases $J=1$ and the remaining parameters are set to zero.}
	\label{fig:Dispersion_d_no_d}
\end{figure}

Returning to our two representative examples (Fig.~\ref{fig:Dispersion_d_no_d}(a,c)), we see that the $T_{2g}$ ground state with $K>0$ retains the characteristic dispersionless mode between $\Gamma$ and $X$ while the $T_{1g}$ state with $K<0$ breaks the continuous degeneracy due to the induced ferromagnetic component resulting in a nearly dispersionless mode. Importantly, in both situations, each of the three spin-wave modes are degenerate along the two paths through the Brillouin zone. In addition, all three modes are degenerate at both the $M/M'$ and $R/R'$ points. These degeneracies are lifted as soon as a small DM term of the form detailed in Eq.~\ref{eq:samplecouplings} is introduced, leading to a path-dependent spectrum, a gapped spin-wave spectrum at $M/M'$ and $R/R'$ as well as mode crossings that are not confined to the BZ boundaries.
In the interest of exploring possible ways to measure this DM induced effect, we provide calculations of inelastic neutron scattering intensities. Again following Refs.~\cite{Toth2015,Petit2011}, we calculate the perpendicular component of the $\left(3\times3\right)$ dynamical spin-spin correlation matrix,
\begin{equation}
\small
\mathcal{S}(\vb{k},\omega) = \sum_{\alpha\beta}\left(\delta_{\alpha\beta}-\frac{q_{\alpha}q_{\beta}}{q^2}\right)\mathcal{S}_{\alpha\beta}\left(\vb{k},\omega\right)
\end{equation}
where
\begin{equation}
\small
\mathcal{S}_{\alpha\beta}(\vb{k},\omega) = \frac{1}{2\pi N}\sum_{mi,nj}e^{i\vb{k}(\vb{r}_{mi}-\vb{r}_{nj})}\int_{-\infty}^{\infty}d\tau e^{-i\omega\tau}\left<S_{mi}^\alpha S_{nj}^\beta\right>,
\end{equation}
$\alpha, \beta \in \left(x,y,z\right)$, $m, n$ are unit cell indices, $N=3$, and $i, j$ again label sublattices. 

\begin{figure}[h]
	\includegraphics[width=1.\columnwidth]{{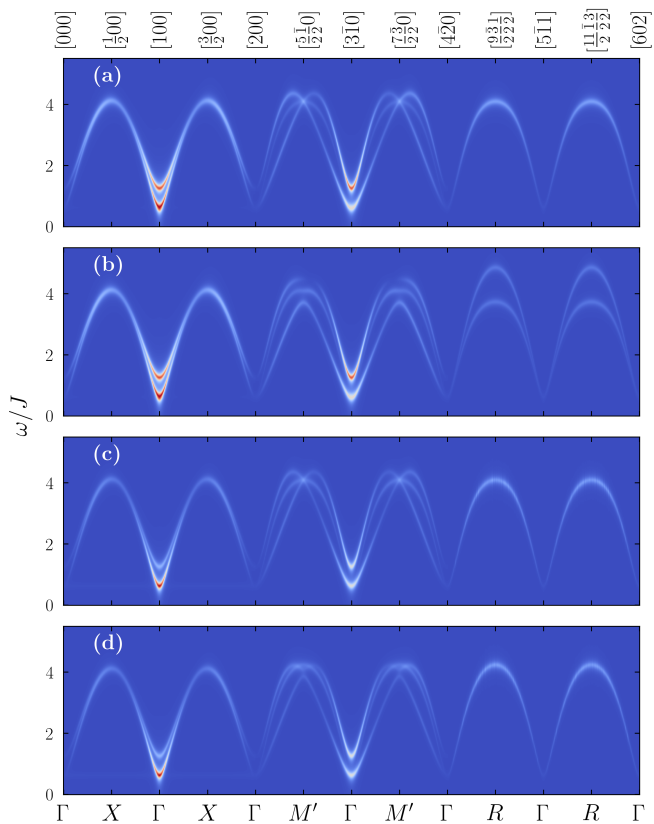}}
	\caption{Relative magnitudes of the dynamical correlation function $\mathcal{S}(\vb{k},\omega)$ for (a) the $T_{1g}$ configuration with $K=-0.1,D=0$, (b) the $T_{1g}$ configuration with $K=-0.1,D=-0.2$, (c) the $T_{2g}$ configuration with $K=0.1,D=0$ and (d) the $T_{2g}$ configuration with $K=0.1,D=-0.2$.}
	\label{fig:INS_T1g_T2g}
\end{figure}

\begin{figure}[h]
	\centering
	\includegraphics[width=0.65\columnwidth]{{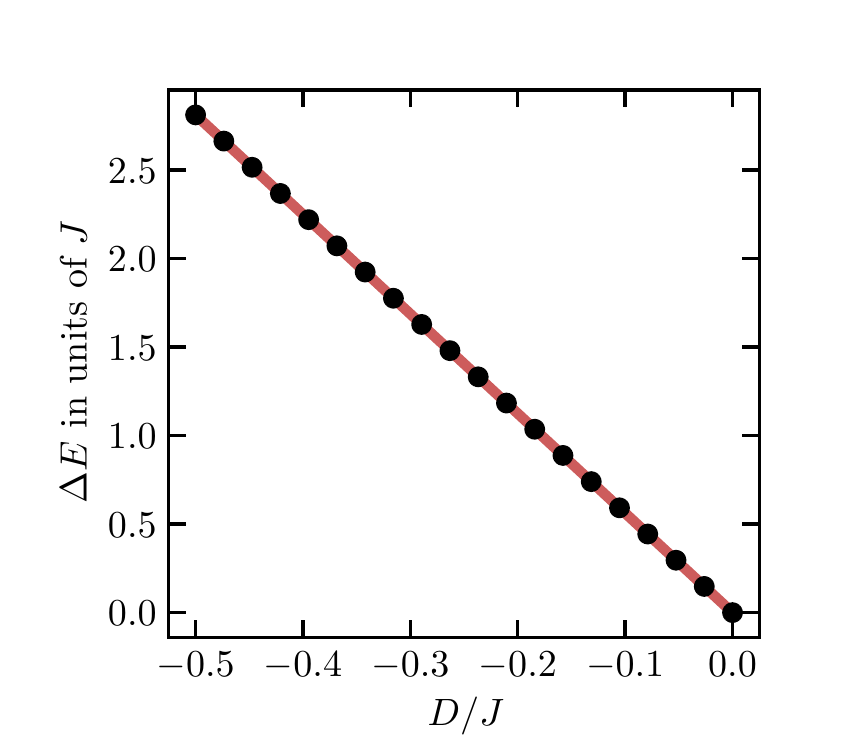}}
	\caption{The difference in $\omega$ of the two visible spin wave modes for the T1 structure at the $R$ point as a function of $D$.}
	\label{fig:Linear_DeltaE}
\end{figure}

We do this calculation for a path containing $M'$ and $R$ since these are effectively the local $x$ and $z$ directions within the coplanar/near-coplanar kagome planes. The path we follow highlights markedly higher intensity in the second Brillouin zones consistent with the elastic neutron scattering intensity appearing for $h,k,\ell$ all even and (nearly) vanishing for $h,k,\ell$ all odd as was reported in~\cite{LeBlanc2014}. As can be seen in Fig.~\ref{fig:INS_T1g_T2g}, the effect of the DM interaction in the $T_{1g}$ configuration is qualitatively obvious since both modes with appreciable intensity at the $R$ point become non-degenerate. Alternatively, for the $T_{2g}$ state, the two modes with appreciable intensity remain degenerate along the path $\Gamma-R$ making the effect of the DM interaction less obvious. Focusing on the $M'$ point, the influence of the DM interaction can again be observed with the effect again far more obvious for the $T_{1g}$ configuration. Nonetheless, the influence of the DM interaction should be measurable at $M'$ in both configurations since all three spin wave modes have appreciable intensity in both cases.
With the effect of the DM interaction being so prevalent in the slice $\mathcal{S}\left(\vb{k}=R,\omega\right)$, we think it useful to investigate the dependence on the splitting of the two visible modes as a function of the DM strength. We define $\Delta E$ as the difference in energy of the top and intermediate modes of the $T_{1g}$ structure at the $R$-point. Figure~\ref{fig:Linear_DeltaE} shows the relationship between $\Delta E$ and $D$ for $D\in\left(-0.5,0\right)$ which is linear, outlining a potential means of comparing the relative DM strength in different T1-ordered materials. We also note the relevance of this effect as it pertains to certain optical measurements such as the work carried out in Ref.~\cite{RodrguezSurez2021}.

\begin{figure}[h]
	\includegraphics[width=1.\columnwidth]{{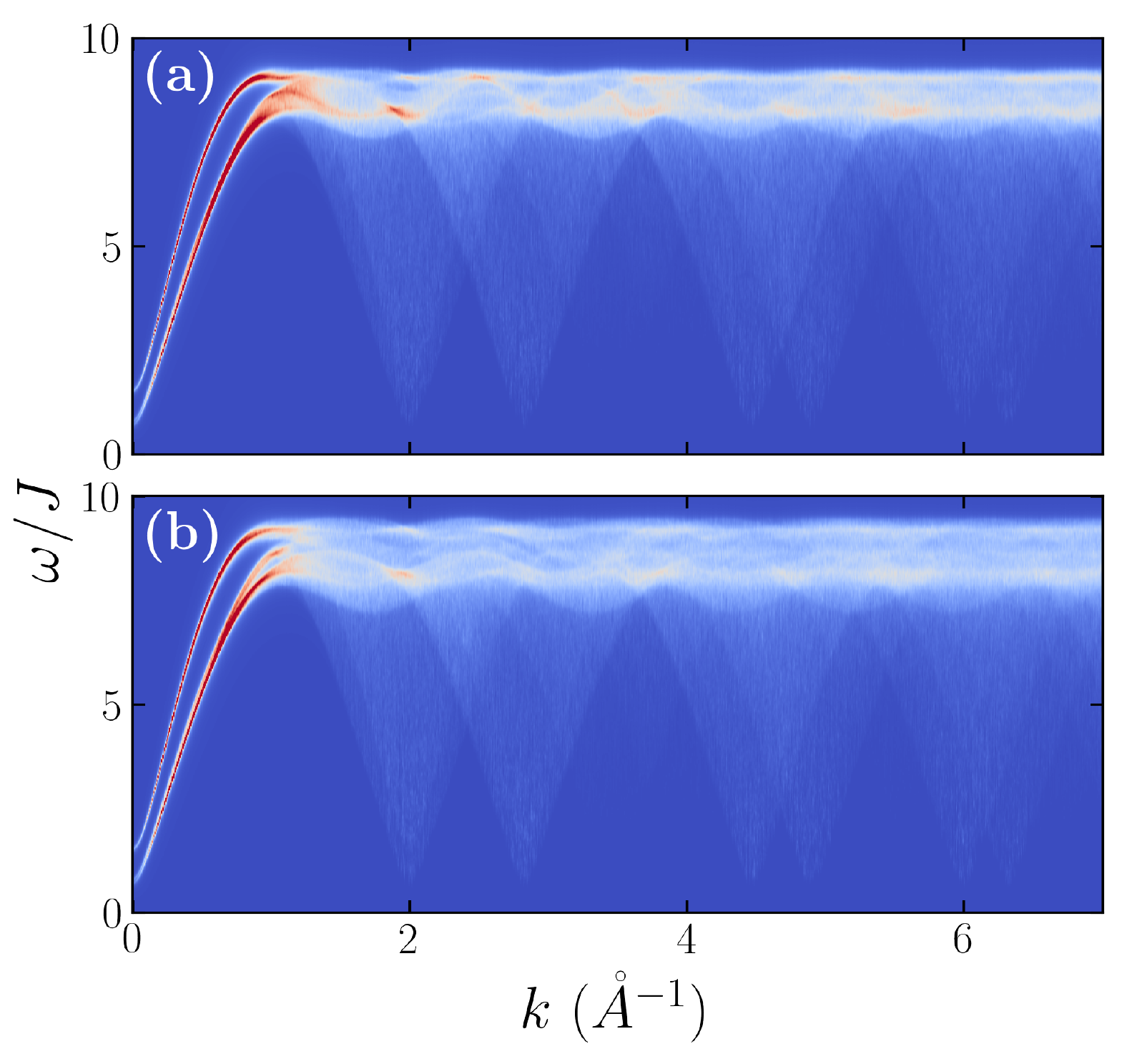}}
	\caption{Comparison of calculated powder INS intensity profiles for (a) the $J_1, J_2, J_3, J_4$ model with $D=0$ and (b) the $J_1, J_2, J_3, J_4$ model with $D=-0.2$.}
	\label{fig:INS_Powder}
\end{figure}

Though single crystal INS data for either the \ce{Mn3X} or the \ce{Mn3AB} systems does not exist to our knowledge, experiments have been conducted on polycrystalline samples of the \ce{Cu3Au}-type \ce{Mn3Ir}~\cite{LeBlanc2021}. In the interpretation of this data, the authors made use of not just nearest-neighbor isotropic exchange $J$ and effective SIA $K_{\text{eff}}$, but also included further neighbor isotropic exchange interactions reaching out to the fourth nearest-neighbor Mn ions. The basis of including these interactions and the initial guesses of their various strengths were owing to a density functional theory (DFT) study by Szunyogh \textit{et al.}~\cite{Szunyogh2009} which predicted alternating ferro/antiferromagnetic couplings, $J_1,J_2,J_3$ and $J_4$ along with a value for $K_{\text{eff}}$. The authors fit this five-parameter model to the powder INS data and found discrepancies with the values predicted by density functional theory (DFT). We wish to point out here that the influence of the DM interaction could substantively alter the powder INS calculations even when its magnitude is relatively weak and should be considered when interpreting the spin-wave spectra in this material.To highlight this, in Fig.~\ref{fig:INS_Powder} we show calculations of the powder INS spectrum for two situations. The first is the case with all values set to the DFT predictions (normalized so that $J_1=1$) while the second includes the same values for $J_1,J_2,J_3,J_4$ and $K_{\text{eff}}$ but also includes DM coupling with value $D=-0.2$. Importantly, the presence of the DM interaction affects not only the total magnon bandwidth, causing higher energy excitations, but also affects the width of the high energy parallel ``tails", increasing this width in $E$ from $\approx 1J$ to $\approx 1.4J$ as can be seen in Fig.~\ref{fig:INS_Powder}. 
\section{Conclusions}

To summarize, we have conducted a symmetry analysis of $L1_2$ \ce{Mn3X} and \ce{Mn3AB} compounds to construct a classical magnetic Hamiltonian which includes terms due to NN and NNN isotropic exchange, NN Kitaev-type anisotropic exchange, NN Gamma-type anisotropic exchange, NN DM interaction and single-ion anisotropy. The magnetic ground states we predict using our model are consistent with previous results obtained considering only NN/NNN isotropic exchange and single-ion anisotropy with the off-diagonal $\mathcal{H}_D$ and $\mathcal{H}_B$ terms making no energetic contribution in large swaths of the ground-state phase diagrams. We show, however, that these terms can strongly influence the elementary excitations of these spin-structures. Reduction in the Brillouin zone symmetry leads to path-dependent spin-wave dispersion in a similar manner to the AB-stacked system as described in~\cite{Zelenskiy2021}. Additionally, gaps are formed at the Brillouin zone boundaries which have linear dependence on the presupposed DM interaction's magnitude. Our inelastic neutron scattering calculations show that both effects should be measurable and highlight a route for quantifying and comparing the relative influence of these cloaked interactions in different \ce{Mn3X} and \ce{Mn3AB} compounds.

\section{Acknowledgments} 
The authors would like to thank D.~Kalliecharan, B.~D.~MacNeil, and C.~D.~Rudderham for their valuable discussions and insights. This work was supported by the Natural Sciences and Engineering Research Council of Canada (NSERC).

\balance
\bibliographystyle{apsrev4-2}
\bibliography{McCoombs_Ref}
\appendix
\section{Symmetry Operations}
In section II we made use of the generating operations of point group $O_h$. These operations can be found in~\cite{Aroyo2006}, but we detail their application to our system here. To apply the generating operations to our coupling matrix $\mathcal{J}_{ij}\left(\vb{r}-\vb{r}^{\prime}\right)$ we must first understand the spin permutations that result from each operation. The full transformation $\mathcal{T}$ associated with each generating operation $g$ is the tensor product between the permutation matrix $\mathcal{P}$ and the associated axial-vector rotation matrix $\mathcal{R}$:
\begin{equation}
	\mathcal{T}(g) = \mathcal{P}(g)\otimes\mathcal{R}(g).
\end{equation}
These permutation and rotation matrices for the six-spin basis used in section II are as follows:
\begin{equation}
\begin{aligned}
\mathcal{P}(C_2) &= \displaystyle \left[\begin{matrix}
1 & 0 & 0 & 0 & 0 & 0\\
0 & 0 & 0 & 0 & 1 & 0\\
0 & 0 & 0 & 0 & 0 & 1\\
0 & 0 & 0 & 1 & 0 & 0\\
0 & 1 & 0 & 0 & 0 & 0\\
0 & 0 & 1 & 0 & 0 & 0
\end{matrix}\right],
\mathcal{R}(C_2) = \displaystyle \left[\begin{matrix}
-1 & 0 & 0\\
0 & -1 & 0\\
0 & 0 & 1
\end{matrix}\right]\\
\mathcal{P}(C_2^{\prime}) &= \displaystyle \left[\begin{matrix}
0 & 0 & 0 & 1 & 0 & 0\\
0 & 0 & 0 & 0 & 1 & 0\\
0 & 0 & 1 & 0 & 0 & 0\\
1 & 0 & 0 & 0 & 0 & 0\\
0 & 1 & 0 & 0 & 0 & 0\\
0 & 0 & 0 & 0 & 0 & 1
\end{matrix}\right],
\mathcal{R}(C_2^{\prime}) = \displaystyle \left[\begin{matrix}
-1 & 0 & 0\\
0 & 1 & 0\\
0 & 0 & -1
\end{matrix}\right]\\
\mathcal{P}(C_3) &= \displaystyle \left[\begin{matrix}
0 & 0 & 1 & 0 & 0 & 0\\
1 & 0 & 0 & 0 & 0 & 0\\
0 & 1 & 0 & 0 & 0 & 0\\
0 & 0 & 0 & 0 & 0 & 1\\
0 & 0 & 0 & 1 & 0 & 0\\
0 & 0 & 0 & 0 & 1 & 0
\end{matrix}\right],
\mathcal{R}(C_3) = \displaystyle \left[\begin{matrix}
0 & 0 & 1\\
1 & 0 & 0\\
0 & 1 & 0
\end{matrix}\right]\\
\mathcal{P}(C_2^{\prime\prime}) &= \displaystyle \left[\begin{matrix}
0 & 0 & 0 & 1 & 0 & 0\\
0 & 0 & 1 & 0 & 0 & 0\\
0 & 1 & 0 & 0 & 0 & 0\\
1 & 0 & 0 & 0 & 0 & 0\\
0 & 0 & 0 & 0 & 0 & 1\\
0 & 0 & 0 & 0 & 1 & 0
\end{matrix}\right],
\mathcal{R}(C_2^{\prime\prime}) = \displaystyle \left[\begin{matrix}
0 & 1 & 0\\
1 & 0 & 0\\
0 & 0 & -1
\end{matrix}\right]\\
\mathcal{P}(\mathcal{I}) &= \displaystyle \left[\begin{matrix}
0 & 0 & 0 & 1 & 0 & 0\\
0 & 0 & 0 & 0 & 1 & 0\\
0 & 0 & 0 & 0 & 0 & 1\\
1 & 0 & 0 & 0 & 0 & 0\\
0 & 1 & 0 & 0 & 0 & 0\\
0 & 0 & 1 & 0 & 0 & 0
\end{matrix}\right],
\mathcal{R}(\mathcal{I}) = \displaystyle \left[\begin{matrix}
1 & 0 & 0\\
0 & 1 & 0\\
0 & 0 & 1
\end{matrix}\right]\\.
\label{eq:generators}
\end{aligned}
\end{equation}
In section IV we made use of representative symmetry operations from each class of point group $O_h$. The character table for $O_h$ is as follows:
\begin{table}[H]
	\centering
	\begin{ruledtabular}
	\begin{tabular}{c|cccccccccc}
		$O_h$ & $E$ & $8C_3$ & $6C_2$ & $6C_4$ & $3C_2^\prime$ & $\mathcal{I}$ & $6S_4$ & $8S_6$ & $3\sigma_h$ & $6\sigma_d$\\ \hline    
		$A_{1g}$ & $+1$ & $+1$ & $+1$ & $+1$ & $+1$ & $+1$ & $+1$ & $+1$ & $+1$ & $+1$ \\ 
		$A_{2g}$ & $+1$ & $+1$ & $-1$ & $-1$ & $+1$ & $+1$ & $-1$ & $+1$ & $+1$ & $-1$ \\
		$E_{g}$ & $+2$ & $-1$ & $0$ & $0$ & $+2$ & $+2$ & $0$ & $-1$ & $+2$ & $0$ \\
		$T_{1g}$ & $+3$ & $0$ & $-1$ & $+1$ & $-1$ & $+3$ & $+1$ & $0$ & $-1$ & $-1$ \\ 
		$T_{2g}$ & $+3$ & $0$ & $+1$ & $-1$ & $-1$ & $+3$ & $-1$ & $0$ & $-1$ & $+1$ \\   
		$A_{1u}$ & $+1$ & $+1$ & $+1$ & $+1$ & $+1$ & $-1$ & $-1$ & $-1$ & $-1$ & $-1$ \\
		$A_{2u}$ & $+1$ & $+1$ & $-1$ & $-1$ & $+1$ & $-1$ & $+1$ & $-1$ & $-1$ & $+1$ \\
		$E_{u}$ & $+2$ & $-1$ & $0$ & $0$ & $+2$ & $-2$ & $0$ & $+1$ & $-2$ & $0$ \\
		$T_{1u}$ & $+3$ & $0$ & $-1$ & $+1$ & $-1$ & $-3$ & $-1$ & $0$ & $+1$ & $+1$ \\
		$T_{2u}$ & $+3$ & $0$ & $+1$ & $-1$ & $-1$ & $-3$ & $+1$ & $0$ & $+1$ & $-1$         
	\end{tabular}%
	\end{ruledtabular}
\caption{Character table for point group $O_h$.}
\label{Tab:CharacterTable} 
\end{table}
The permutation and rotation matrices associated with the representative operations $g$ from each class are as follows:
\begin{equation}
\begin{aligned}
\mathcal{P}(C_3) &= \displaystyle \left[\begin{matrix}0 & 0 & 1\\1 & 0 & 0\\0 & 1 & 0\end{matrix}\right], 
\mathcal{R}(C_3) = \displaystyle \left[\begin{matrix}0 & 0 & 1\\1 & 0 & 0\\0 & 1 & 0\end{matrix}\right]\\
\mathcal{P}(C_2) &= \displaystyle \left[\begin{matrix}1 & 0 & 0\\0 & 0 & 1\\0 & 1 & 0\end{matrix}\right],
\mathcal{R}(C_2) = \displaystyle \left[\begin{matrix}0 & 1 & 0\\1 & 0 & 0\\0 & 0 & -1\end{matrix}\right]\\
\mathcal{P}(C_4) &= \displaystyle \left[\begin{matrix}1 & 0 & 0\\0 & 0 & 1\\0 & 1 & 0\end{matrix}\right],
\mathcal{R}(C_4) = \displaystyle \left[\begin{matrix}0 & -1 & 0\\1 & 0 & 0\\0 & 0 & 1\end{matrix}\right]\\
\mathcal{P}(C_2^\prime) &= \displaystyle \left[\begin{matrix}1 & 0 & 0\\0 & 1 & 0\\0 & 0 & 1\end{matrix}\right],
\mathcal{R}(C_2^\prime) = \displaystyle \left[\begin{matrix}-1 & 0 & 0\\0 & -1 & 0\\0 & 0 & 1\end{matrix}\right]\\
\mathcal{P}(\mathcal{I}) &= \displaystyle \left[\begin{matrix}1 & 0 & 0\\0 & 1 & 0\\0 & 0 & 1\end{matrix}\right],
\mathcal{R}(\mathcal{I}) = \displaystyle \left[\begin{matrix}1 & 0 & 0\\0 & 1 & 0\\0 & 0 & 1\end{matrix}\right]\\
\mathcal{P}(S_4) &= \displaystyle \left[\begin{matrix}1 & 0 & 0\\0 & 0 & 1\\0 & 1 & 0\end{matrix}\right],
\mathcal{R}(S_4) = \displaystyle \left[\begin{matrix}0 & -1 & 0\\1 & 0 & 0\\0 & 0 & 1\end{matrix}\right]\\
\mathcal{P}(S_6) &= \displaystyle \left[\begin{matrix}0 & 0 & 1\\1 & 0 & 0\\0 & 1 & 0\end{matrix}\right],
\mathcal{R}(S_6) = \displaystyle \left[\begin{matrix}0 & 0 & 1\\1 & 0 & 0\\0 & 1 & 0\end{matrix}\right]\\
\mathcal{P}(\sigma_h) &= \displaystyle \left[\begin{matrix}1 & 0 & 0\\0 & 1 & 0\\0 & 0 & 1\end{matrix}\right],
\mathcal{R}(\sigma_h) = \displaystyle \left[\begin{matrix}-1 & 0 & 0\\0 & -1 & 0\\0 & 0 & 1\end{matrix}\right]\\
\mathcal{P}(\sigma_d) &= \displaystyle \left[\begin{matrix}1 & 0 & 0\\0 & 0 & 1\\0 & 1 & 0\end{matrix}\right],
\mathcal{R}(\sigma_d) = \displaystyle \left[\begin{matrix}0 & 1 & 0\\1 & 0 & 0\\0 & 0 & -1\end{matrix}\right]\\.
\label{eq:representatives}
\end{aligned}
\end{equation}
with $\mathcal{P}(g)$ in the $[\vb{S}_1,\vb{S}_2,\vb{S}_3]$ basis and $\mathcal{R}(g)$ in the $[S_i^x,S_i^y,S_i^z]$ basis.

\section{Unit vectors}

In section II, the symmetry-allowed magnetic Hamiltonian was defined as~\ref{eq:maghamterms}. The various unit vectors, $\hat{\vb{D}}_{ij}, \hat{\vb{n}}_i, \hat{\vb{n}}_{ij}, \hat{\vb{l}}_{ij}, \hat{\vb{m}}_{ij}$, used to define the couplings are as follows:
\begin{equation}
\begin{aligned}
\vb{D}_{12}=\vb{D}_{1'2'}=D\hat{\vb{y}} \\
\vb{D}_{1'2}=\vb{D}_{12'}=-D\hat{\vb{y}} \\
\vb{D}_{31}=\vb{D}_{3'1'}=D\hat{\vb{x}} \\
\vb{D}_{31'}=\vb{D}_{3'1}=-D\hat{\vb{x}} \\
\vb{D}_{23}=\vb{D}_{2'3'}=D\hat{\vb{z}} \\
\vb{D}_{23'}=\vb{D}_{2'3}=-D\hat{\vb{z}} \\
\hat{\vb{n}}_1 = \hat{\vb{n}}_{1'} = \hat{\vb{z}} \\
\hat{\vb{n}}_2 = \hat{\vb{n}}_{2'} = \hat{\vb{x}} \\
\hat{\vb{n}}_3 = \hat{\vb{n}}_{3'} = \hat{\vb{y}} \\
\hat{\vb{n}}_{12} = \hat{\vb{n}}_{12'} = \hat{\vb{n}}_{21'} = \hat{\vb{n}}_{1'2'} = \hat{\vb{y}} \\
\hat{\vb{n}}_{13} = \hat{\vb{n}}_{13'} = \hat{\vb{n}}_{31'} = \hat{\vb{n}}_{1'3'} = \hat{\vb{x}} \\
\hat{\vb{n}}_{23} = \hat{\vb{n}}_{23'} = \hat{\vb{n}}_{32'} = \hat{\vb{n}}_{2'3'} = \hat{\vb{z}} \\
\hat{\vb{l}}_{12} = \hat{\vb{l}}_{12'} = \hat{\vb{l}}_{21'} = \hat{\vb{l}}_{1'2'} = \hat{\vb{x}} \\
\hat{\vb{l}}_{13} = \hat{\vb{l}}_{13'} = \hat{\vb{l}}_{31'} = \hat{\vb{l}}_{1'3'} = \hat{\vb{y}} \\
\hat{\vb{l}}_{23} = \hat{\vb{l}}_{23'} = \hat{\vb{l}}_{32'} = \hat{\vb{l}}_{2'3'} = \hat{\vb{x}} \\
\hat{\vb{m}}_{12} = \hat{\vb{m}}_{12'} = \hat{\vb{m}}_{21'} = \hat{\vb{m}}_{1'2'} = \hat{\vb{z}} \\
\hat{\vb{m}}_{13} = \hat{\vb{m}}_{13'} = \hat{\vb{m}}_{31'} = \hat{\vb{m}}_{1'3'} = \hat{\vb{z}} \\
\hat{\vb{m}}_{23} = \hat{\vb{m}}_{23'} = \hat{\vb{m}}_{32'} = \hat{\vb{m}}_{2'3'} = \hat{\vb{y}}.
\label{eq:unit_vectors}
\end{aligned}
\end{equation}

\end{document}